\documentclass[nonacm,sigconf,10pt]{acmart}

\usepackage[english]{babel}
\usepackage{blindtext}
\usepackage{enumitem}
\usepackage{duckuments}

\renewcommand\footnotetextcopyrightpermission[1]{} 
\setcopyright{none}

\usepackage{dsfont}
\usepackage{mathrsfs,amsmath}
\usepackage{kantlipsum}
\usepackage{xcolor}
\usepackage[normalem]{ulem}
\usepackage{graphicx}
\usepackage[linesnumbered,ruled,noline]{algorithm2e}
\usepackage{subcaption}

\usepackage{tikz}
\newcommand*\circled[1]{\tikz[baseline=(char.base)]{
            \node[shape=circle,fill,inner sep=0.7pt] (char) {\textcolor{white}{#1}};}}
\newcommand{\oc}[1]{\textcolor{orange}{OC: #1}}

\newcommand{\name}{MapMesh\xspace}

\SetKwInput{KwInput}{Input}                
\SetKwInput{KwOutput}{Output}              
\SetKwBlock{Loop}{Loop}{}
\SetKwBlock{Func}{Function}{}
\SetKwBlock{While}{While}{}
\SetKwBlock{If}{If}{}
\SetKwBlock{Else}{Else}{}
\newcommand{\nonl}{\renewcommand{\nl}{\let\nl\oldnl}}

\newcommand\blfootnote[1]{
    \begingroup
    \renewcommand\thefootnote{}\footnote{#1}
    \addtocounter{footnote}{-1}
    \endgroup
} 

\acmDOI{}

\acmISBN{}

\acmPrice{}

\usepackage{tikz}
\usepackage{enumitem}

\begin{document}
\title[]{Scalable Routing in a City-Scale Wi-Fi Network for\\Disaster Recovery}

\author{\large Ziqian Liu*, Om Chabra*, James Lynch, Chenning Li, Manya Ghobadi, and Hari Balakrishnan\\MIT CSAIL}\vspace{2mm}


\renewcommand{\shortauthors}{}

\begin{abstract}
    In this paper, we present a new city-scale decentralized mesh network system suited for disaster recovery and emergencies. When wide-area connectivity is unavailable or significantly degraded, our system, \name, enables static access points and mobile devices equipped with Wi-Fi in a city to route packets via each other for intra-city connectivity and to/from any nodes that might have Internet access, e.g., via satellite. The chief contribution of our work is a new routing protocol that scales to millions of nodes, a significant improvement over prior work on wireless mesh and mobile ad hoc networks. Our approach uses detailed information about buildings from widely available maps---data that was unavailable at scale over a decade ago, but is widely available now---to compute paths in a scalable way. 
\end{abstract}

\maketitle
\section{Introduction}
\label{s:intro}

\blfootnote{*Equal contribution}
This paper describes a wireless routing protocol that scales well to many millions of Wi-Fi devices, both static and mobile, in a dense urban environment. We show that by using accurate geospatial maps of buildings in urban areas---information now available for most urban areas around the world---we can route packets for millions of devices in over 650,000 buildings in an urban area such as New York while storing only a few hundred routing entries. 
To our knowledge, and in our simulation results, we show that these results are orders of magnitude more scalable than prior methods developed in the long line of research on mobile and wireless mesh networks to date (\S\ref{sec:related_work}).

Before describing how we achieve these results and justifying the underlying assumptions, the natural question in the reader's mind is likely to be ``why does this matter?'' and ``isn't this already a solved problem?'' We will address the second question at length in the next section---remarking here only that no prior work has demonstrated scalability to over a few thousand nodes---and focus instead on the first question to motivate this paper.

As noted in a recent HotNets paper~\cite{citymesh}, the Internet has become extremely centralized both at the network and application layers thanks to Internet Service Provider consolidation, extensive amounts of physical co-location, and cloud computing as the primary way to deploy application software. As a result, our networks and applications are vulnerable to disasters both natural and human, and attacks both digital and physical.

When a disaster strikes, access to the broader Internet may be fully or partially destroyed or may be severely hampered. Entire regions may lose connectivity, or only a small fraction of nodes may have connectivity via technologies such as satellite networks. In all these cases, if it were possible to provide communication capabilities {\em within a region}, one could support in-region applications such as status updates, messaging, simple financial transactions and commerce, regional emergency and safety services, and connectivity to nodes equipped with Internet access. This network would provide lower bandwidth than usual but could be adequate for these applications during times of duress. The HotNets paper mentioned above made the case for such {\em decentralized fallback networks} (DFNs). 

This paper is inspired by that vision. We develop a complete solution for a city-scale DFN that uses Wi-Fi devices to achieve connectivity. Unlike in the HotNets paper, we support not only Wi-Fi access points (APs) but also mobile devices such as smartphones in the system. And while the HotNets paper showed how maps could be used, the solution there had excessive transmission overhead (each successful packet required 15-20$\times$ as many transmissions as in an optimal unicast scheme, and in addition required hundreds of bytes of packet header to send even small payloads. In short, it was unscalable, left numerous key details open, and did not demonstrate any definitive results. 

The primary focus of this paper is on scalable wireless routing for DFN applications. To achieve scalability, our proposed scheme eliminates all probe, control, and routing messages between devices. It aggressively uses map data, which is now easily available to route packets between devices by developing a way to route packets across a sequence of buildings to go from any source to the destination. Devices need to know which building they are in, but bootstrap all other knowledge by overhearing other packet broadcasts during operation. The reason why this approach works is because the vast majority of Wi-Fi devices are in or near buildings most of the time (e.g., Wi-Fi access points and most people).\footnote{This network design does not support users who are not close to a building at this time.}

We show how to compute good paths between buildings, how to compress these paths using waypoints that encode turns and departures from linear paths, how to structure a simple grid-based building addressing scheme in any urban area, and how to suppress transmissions to avoid the high overheads in prior work. 

Because of our use of urban maps of buildings, we call our system \name. We have implemented \name on a small campus testbed and in ns-3 and have conducted several simulations across several cities in the world. Our key contributions and findings are:

\begin{itemize}
    \item To our knowledge, \name is the first building map-based routing scheme for wireless and mobile device networks.

    \item We have simulated \name in 60 cities across 40 countries and 6 continents. The number of devices varies from 762 (Vatican City) to 909,010 (New York City) with an average one-way packet loss rate varying from 0 to 80\%. We achieve a delivery success rate between 20 to 95 percentage points higher than multiple versions of GPSR, the best prior scheme for scalable ad hoc routing. 
    
    
    \item In terms of required memory, we find that in cities with hundreds of thousands of buildings, we require a routing table size of a few hundred to a few thousand entries (a few kilobytes). 
    %
    
    
    \item When we consider packet transmission (bandwidth) overheads, \name is between 3$\times$ and 7$\times$ superior to prior schemes because it uses no routing messages during its operation.
\end{itemize}

This work does not raise ethical issues.

\section{Related Work}
\label{sec:related_work}

Prior work on ad hoc and wireless mesh routing spans many dozens of papers~\cite{kim_geographic_nodate, dream, gpsr, lar, ilar, corson_1995, tbrpf, cedar, qolsr, alarm, ant-dymo, rahul_geographical}, and may be broadly classified into two categories: \textbf{topological routing}, which send different forms of control packets to determine routes, and \textbf{geographic routing}, which rely on out-of-band location information~\cite{manet_taxonomy}.



\noindent
\textbf{Topological routing.} A network is topological if nodes has no information other than a non-physical address. For these networks, since no other information is given, nodes rely on sending control packets between each other to gain information about the network. A source needs to send some form of a destination discovery packet to first locate the destination and then use the information gathered from these control packets to determine where to forward packets. Topological schemes fall into one of two categories: {\em proactive} and {\em reactive}. 

In proactive protocols~\cite{dsdv, batman, batman_adv, kiran_experimental_2018, olsr, star, tbrpf, wrp, rip}, every node continually updates its routing table by periodically (and perhaps partially) flooding the network with metadata packets. For every packet from a source to a destination, intermediate nodes will forward the packet by matching the destination address with the information they obtained from previously transmitted/received metadata packets. However, due to flooding, these proactive protocols are inherently limited in their scalability. The constant sending of metadata packets to maintain such a large state becomes cumbersome.

By contrast, reactive protocols~\cite{corson_1995, park_1997, royer2000implementation, aodv, tora, aqor, dymo, ara, dsr, odlw, champ} update routing tables only when a source tries to communicate with a destination. However, while incurring less overhead than proactive schemes, obtaining information from reactive messages becomes challenging even at a scale of a few thousand nodes. In particular, every time the network topology changes (a node enters, leaves, or moves), a burst of messages are needed to update information on the destination and routes. As a network scales to many thousands of devices, churn becomes more frequent, requiring nearly the same amount of messaging overhead as the proactive approaches. 

One approach to this problem is exemplified by DYMO~\cite{dymo}, also known as AODVv2, which warns that it is ``best suited for relatively sparse traffic scenarios where any particular router forwards packets to only a small percentage of the AODVv2 routers in the network''. The dense control packet traffic in larger networks will undermine reactive protocols like DYMO. 

There are hybrid approaches~\cite{iwata_scalable_1999, ietf-manet-zone-zrp-04, ant-aodv, zhls, mzrp, lanmar, hopnet, a4lp}, which combine active flooding in a small cluster of nodes while using reactive on-demand routing for inter-cluster communication. However, these approaches only seek to reduce the latency of reactive protocols while reducing the flooding of proactive protocols; hence, they still face the same scalability issues of requiring large amounts of control traffic.


\noindent
\textbf{Geographic routing.} Here, nodes use geographic positions obtained from sensors like GPS to reduce the need for control packets.  During destination discovery, a source will find a destination's geographical position through a system such as the Grid Location Service (GLS)~\cite{li_gls}, which provides information on the location of any destination. In geographic routing, nodes do not need to flood the network with control packets to perform a destination lookup. Instead, packet forwarding occurs in greedy fashion, with complicated mechanisms to overcome voids and dead-ends. 
These approaches largely only send metadata to their one-hop neighbors while some send no metadata packets~\cite{gpsr, dream, goafr, goafr_plus, gdstr, pvex, face_routing, kranakis_compass, shanmuga_gpsr, liangli_gpsr}. One well-known approach, GPSR~\cite{gpsr}, works by having each node store the positions of its one-hop neighbors and forward a packet towards a destination greedily until no further progress can be made. At this point, the protocol enters a recovery mode that uses a deterministic approach to locate a possible path around the void. 

While a combination of GLS \& GPSR is widely believed to be the most scalable solution to creating a MANET, most prior papers (including GPSR) rely on near-perfect geographic information. However, in practice, in indoor settings the best commodity location precision obtainable today (after years of research and practice) has about 15 meters of error (we have gathered this data from high-end smartphones running iOS and Android). Past research shows even a localization error that is 10\% of the radio range can lead to a degradation of end-to-end packet delivery rate even in a network of only 100 nodes~\cite{localization_error}. Additionally, while some prior works do account for this location inaccuracy, they do so by sending additional control packets, which puts a large burden on the network~\cite{rao_geographic_nodate}. We compare \name to GPSR variants in \S\ref{s:eval}.

\noindent
\textbf{Vehicular ad hoc networks (VANETs).}
VANETs, due to their highly dynamic nature, face significant challenges from position inaccuracies when using geographic routing. Previous research such as AGPSR~\cite{silva_adaptive_2018}, MM-GPSR~\cite{yang_improvement_2018}, GPSR-L~\cite{rao_gpsr-l_2008}, AGF~\cite{naumov_evaluation_2006}, and CBF~\cite{fusler_contention-based_nodate} has sought to address this issue by adapting established MANET routing algorithms. For example, GPSR-L~\cite{rao_gpsr-l_2008} introduces the concept of lifetime to the selection of neighbors to account for vehicular movement. However, these solutions do not handle position errors. Some VANETs studies have used road maps for routing. Since vehicles are confined to roads, works like GSR (Geographic Source Routing)~\cite{lochert_routing_2003}, GPCR~\cite{lochert_geographic_2005}, GPSRJ+~\cite{lee_enhanced_2007}, A-STAR~\cite{kanade_-star_2004}, and GyTAR~\cite{jerbi_improved_2007, jerbi_towards_2009} incorporate predictable movements along roadways to guide routing decisions. They reframe geographical routing as a problem of routing between road segments. We use maps differently, not worrying about roads or vehicles but processing building data to create connected paths.

\noindent
\textbf{Static mesh networks.} Nodes here don't move. Most designs~\cite{roofnet, freifunk, afanasyev2008analysis, msr_mesh, brik_measurement_2008} focus on high throughput with metrics like the expected transmission count or time and use topological ad hoc routing (e.g., Srcr in Roofnet) due to the lack of mobility. This approach floods link-state metrics to create good network paths, but none of these prior works has demonstrated scalability beyond several hundred to a few thousand nodes. 
\section{System Design of \name}
A city-scale DFN aims to provide local communications across a region where access to the wide-area Internet is unavailable or significantly hampered.
DFNs mitigate communication outages during man-made and natural disasters by providing a backup network with no single point of failure.
For DFNs at the city-scale, we propose \name.

\name implements a Wi-Fi mesh network with a flat topology that provides a resilient, decentralized communication medium
across a metropolitan area.
It ensures that, even during a major Internet outage, users maintain connectivity through a self-organizing, local wireless network.
This mesh network interconnects both personal mobile devices and stationary access points within the city to each other to establish direct peer-to-peer communication as well as connecting otherwise offline users to specific locations that have working Internet access (for example, a gateway location with Internet access via a satellite ISP).
\name is best suited for applications with low bandwidth requirements that tolerate high latency, enabling applications such as instant messaging, financial transactions, and large-scale dissemination of information like emergency alerts.
The reliable operation of these applications directly impacts safety during disasters.

\subsection{\name Design}
\name's core strength is that it does not require that participants in the network install any new hardware infrastructure.
We design \name for compatibility with existing access points and mobile devices that are already ubiquitous in cities.
Taking advantage of the high density and broad coverage of these devices and their existing capabilities, \name's design requires only software changes to access points and an application for mobile devices, obviating the need for dedicated hardware, making deployment feasible without significant investment in new infrastructure.
By themselves, extant commercial, residential, and public Wi-Fi distribution systems create an expansive network that, when joined together, provides robust, continuous coverage across a city.
Furthermore, since many access points remain operational even during partial network failures -- such as localized power outages or disruptions to backbone connectivity -- \name realizes the resilience properties of a DFN.

In order to scale to millions of nodes, \name implements a flat topology to eliminate central points of failure and ensure all devices are treated equally across the network.
Further, \name simplifies the role of a device participating in the network by requiring only that a device decide whether it should rebroadcast a packet or drop it based on its location.
\name does away with link layer transmissions and frame acknowledgments by relying on packet broadcasts and the redundancy of devices participating in the mesh.

At the core of \name is the concept of a \emph{postbox}, a message storage mechanism that provides users with a persistent destination for messages sent to them.
Similar to the mechanism described in~\cite{citymesh}, postboxes solve the problem of not knowing the physical location of an intended user at the time a message is sent.
Functionally similar to an IMAP email server, a user's postbox serves as a cache of messages to be polled periodically by the user's mobile device when they wish to retrieve new messages.
When a user polls, their mobile device encrypts and attaches its current location so that the postbox knows where to send the requested messages. 

This approach allows users to anchor their communications to specific access points or host devices situated in frequently visited locations, such as residences or workplaces.

Users have the flexibility to designate multiple postboxes, each serving distinct purposes, such as communication within different peer groups or application-specific data storage.
The postbox abstraction is designed to accommodate a variety of data types, enabling applications within the network to store and retrieve information seamlessly.
By integrating these postboxes as an essential component of \name, our system ensures robust, location-stable communication pathways that adapt to normal user movement around the city.

To securely send messages between users over \name, users exchange contact information out of band, either in advance of an outage over the Internet, in person, etc.
This contact information includes both a user's selection of mailbox locations, as well as a public key, generated by the recipient's mobile device, that are used by the sender to encrypt the message before transmitting over the network.
We assume that a user selects postbox locations in advance of an outage.

\begin{figure}[t]
    \centering
    \begin{subfigure}[b]{1.0\linewidth}
    \includegraphics[width=1.0\linewidth]{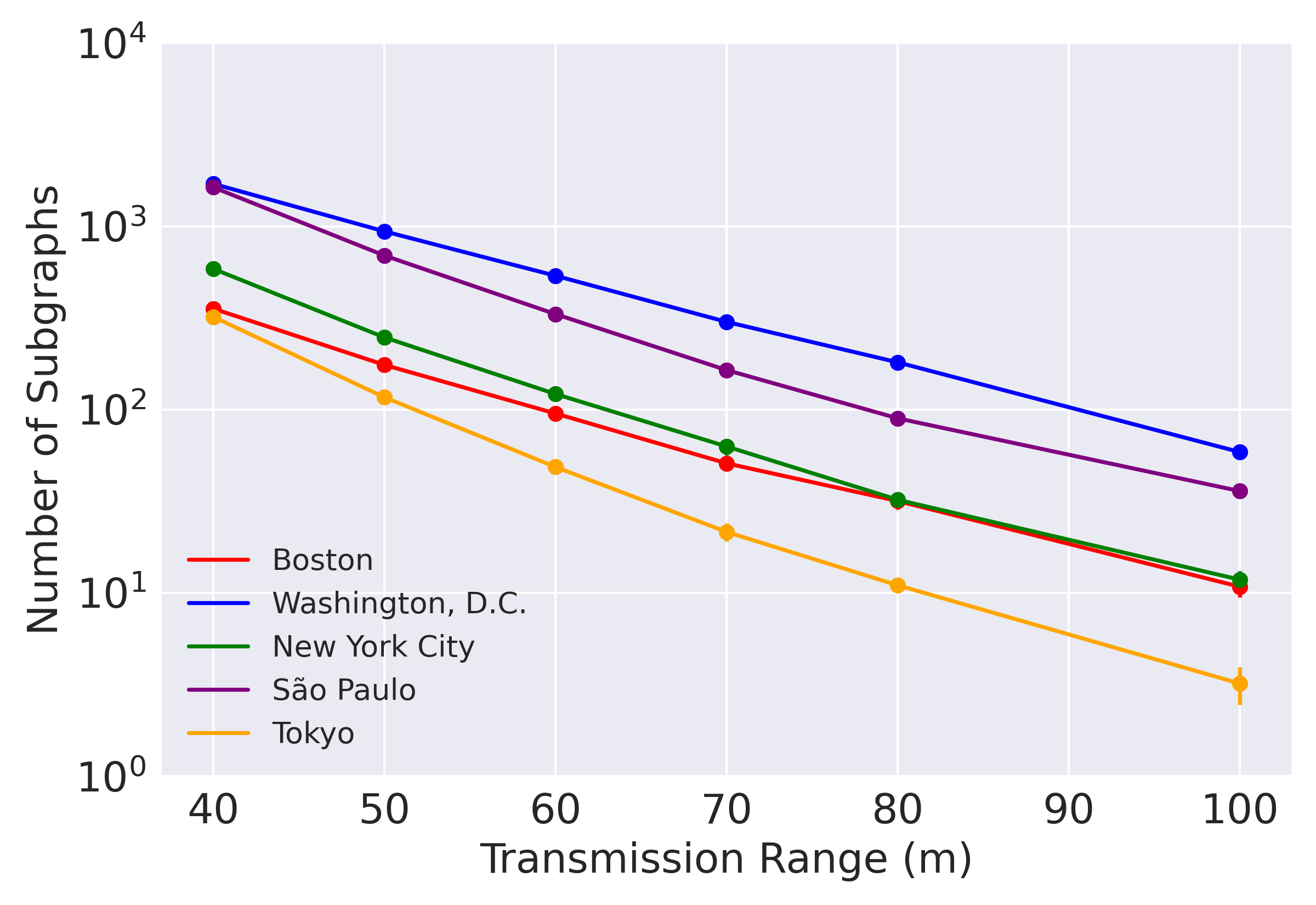}
    \subcaption{Number of subgraphs vs. Wi-Fi transmission range, by city.}
    \label{subfig:subgraphs}
    \end{subfigure}
    \begin{subfigure}[b]{1.0\linewidth}
    \includegraphics[width=1.0\linewidth]{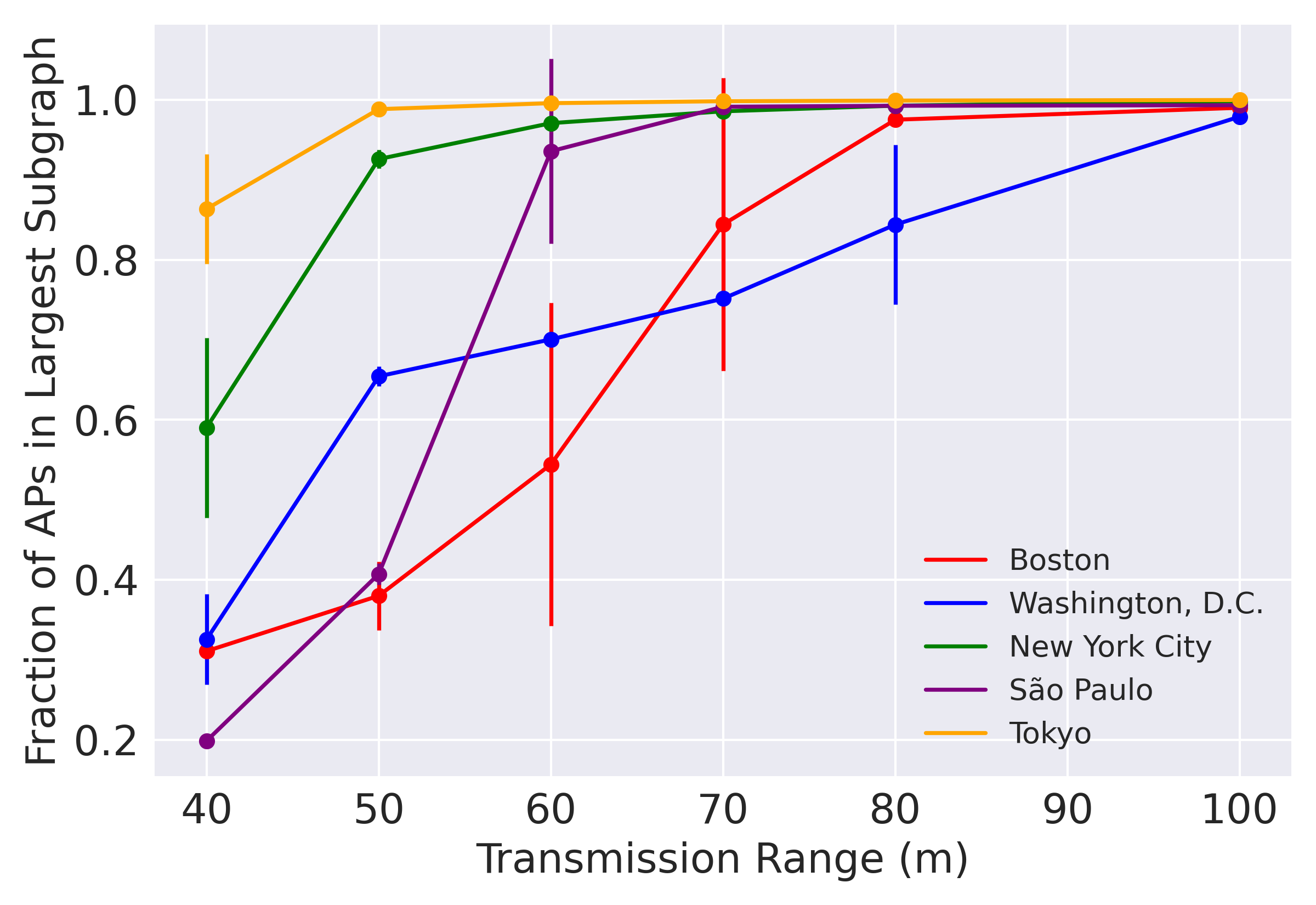}
    \label{subfig:nodes-in-subgraph}
    \subcaption{Fraction of APs in largest connected subgraph vs. Wi-Fi transmission range, by city.}
    \end{subfigure}
    \caption{Impact of transmission range on AP connectivity within a city.}
    \label{fig:apdistribution}
\end{figure}

\subsection{Feasibility of \name}
To evaluate the feasibility of \name, we conduct simulations using real-world urban maps.
High-fidelity map data with detailed building information are freely available through sources like OpenStreetMap~\cite{OpenStreetMap}.
For these simulations, we want to determine whether it is reasonable to claim that a sufficient number of well-placed access points are already present in cities that can connect users across the region.
To do so, we uniformly distribute access points within building footprints across the city as an estimate of existing access point coverage at a set density, here one access point per 200~m$^2$, with at least one access point in each building.
We claim that this density estimate is appropriate, or even conservative, as a city contains many dense commercial wireless distribution systems and residential complexes with dedicated access points for each unit.
For each city we then construct graphs of connectivity between the access points, using a given transmission range as a cutoff.

For any mesh network built using existing access points, the number of devices within the largest subgraph in the city should be as high as possible.
Further, there should be as few ``islands of connectivity'' as possible to maximize connectivity under \name.
Our feasibility results, presented in Figure~\ref{fig:apdistribution}, demonstrate that the majority of access points remain interconnected under reasonable estimates of access point density and wireless communication range but that the extent varies between regions.
This finding underscores the inherent resilience of our approach, suggesting that a \name deployment forms a highly connected network, even without considering the presence of mobile nodes or supplemental hardware infrastructure.



\section{Routing via Buildings}
Routing is a traditionally well-studied problem with rich literature. However, previous work largely considers dealing with routing between each individual device. These protocols often require some out-of-band information (i.e. GPS) for each individual device or transmit control messaging information between each device. To keep routing information updated, these protocols largely require some form of updating or continual messaging as mobile devices continue to move throughout the network. As the number of devices in a network scale, collecting this information, maintaining this information, and relaying this information between devices becomes extremely challenging. All together, when millions of devices are present in a network, handling any information for each separate devices becomes extremely cumbersome and will not scale to practical, realistic deployments. 

Additionally, mobile devices and Wi-Fi access points are highly constrained. Many of these devices have high loss rates with neighboring links. This is worsened by the fact that wireless spectrum is limited, leading to retransmissions of packets wasting crucial limited bandwidth. What's worse, if packets fail to deliver, many higher-layer networking protocols will often retransmit failed packets, causing a dropped packet to re-traverse the network and create more congestion. 

Hence, we design a routing scheme which:

\begin{itemize}
    \item \textit{Requires a small amount of information to be maintained by each device. The content a device stores does not scale directly with the number of devices in the network.}
    \item \textit{Reduces the number of transmissions without compromising on deliverability. Our approach maximizes the likelihood of a packet being received while ensuring that excess transmissions are rare.}
    \item \textit{Is resilient to device failure and churn if a group of devices fails to participate in the network.}
\end{itemize}

In contrast to previous work, we take a new approach that gathers information about a group of devices at once, finding out information about buildings. This gives us two key advantages: (1) buildings are likely to have dense deployments of both Wi-Fi access points and users with mobile devices and (2) we can use building placement data to gather information about neighboring buildings. By using city maps with detailed building information, the precise geographic footprint and area of a building and the number of floors and their heights can be obtained before any network usage. Accurate maps are often collected and maintained by organizations and municipalities using satellites and drones to collect accurate map data~\cite{googleOpenBuilding}. Another unique advantage of building information is that buildings rarely change on the time span of a network being operational. This is in contrast to device-level information that changes often and is subject to numerous external environmental factors.

\subsection{Source Routing}

One approach to using building information to send packets from a source to a destination device is source routing. The source device can use the map of buildings in a city to compute a route (path) expressed as a sequence of buildings, each building being denoted by a unique ID (e.g., from the map database).  The source can specify this sequence in the packet header. Every intermediate node that overhears this packet can decide, using the map (or more simply by filtering on the buildings it is close to), if it should ignore the packet because it is not in or near any building specified in the header, or if it should forward it (or, as we will see later, to consider forwarding it and to suppress the potential transmission if it overhears the transmission from a ``better'' forwarding device).

This approach will likely result in the packet being delivered to the destination as long as a suitable path of buildings is found. However, for this approach to be practical and usable, we must tackle the following problems:

\begin{enumerate}
    \item Given the coordinates and sizes of buildings, what is the best way to compute a building path between source and destination? We address this question in \S\ref{s:rtgmetric} by proposing a simple metric to construct minimum-cost paths from a map of buildings in a city. These paths aim to increase the likelihood of successful packet delivery by avoiding marginal links that may have low packet delivery rates. Traditional mesh routing protocols use probe packets to measure packet delivery and distribute reachability information using these link metrics, but as discussed earlier these methods don't scale to more than a few thousand devices. 
    
    \item As our scheme uses no control messages sent by devices and is unaware of their specific locations (both of these help with scalability), there is no guarantee that there is actually a usable wireless link between two buildings, even if those buildings are on the minimum-cost path. Thus to increase the likelihood of finding a working path, we must not restrict the packet forwarding devices to be within the specified buildings alone, but extend to nearby nodes. We propose a mechanism to tackle this problem in \S\ref{s:conduit} using {\em conduits}, which provide a type of geofence around the chosen path sequence; devices can easily check whether they are within the geofence, and may rebroadcast (forward) the packet only if they are.
    
    \item Encoding a building sequence in a packet header does not scale well across more than a few hundred meters of total path length. We must find a more succinct approach. We develop a mechanism in \S\ref{s:waypoints} to compress a path into a sequence of {\em waypoint buildings}, which represent only places where the chosen path makes turns that depart from a prior linear sequence.
    
    \item Waypoints by themselves don't make packet header lengths independent of the length of the path. Moreover, each source in the network in this approach needs to maintain the entire global map and must compute every route that it needs, posing a non-trivial memory and computing burden on many mobile devices. We tackle this problem in \S\ref{s:gridrtg} with a grid-oriented addressing scheme for buildings and a compact routing table for each device, which depends on its grid position. We also show that grid routes can be precomputed efficiently at a central server and pushed to devices prior to when they need to be used. This solution replaces source routing.
    
    \item Last but not least, this approach will lead to numerous excess transmissions when there are multiple devices in or near the same building that overhear a given transmission and rebroadcast the packet. We describe algorithms in \S\ref{s:suppression} for inter-building and in-building suppression that significantly reduce this overhead.
    
\end{enumerate}




\subsection{Building Graph and Path Selection}
\label{s:rtgmetric}
\begin{figure}[t]
    \centering
    \includegraphics[width=1\linewidth]{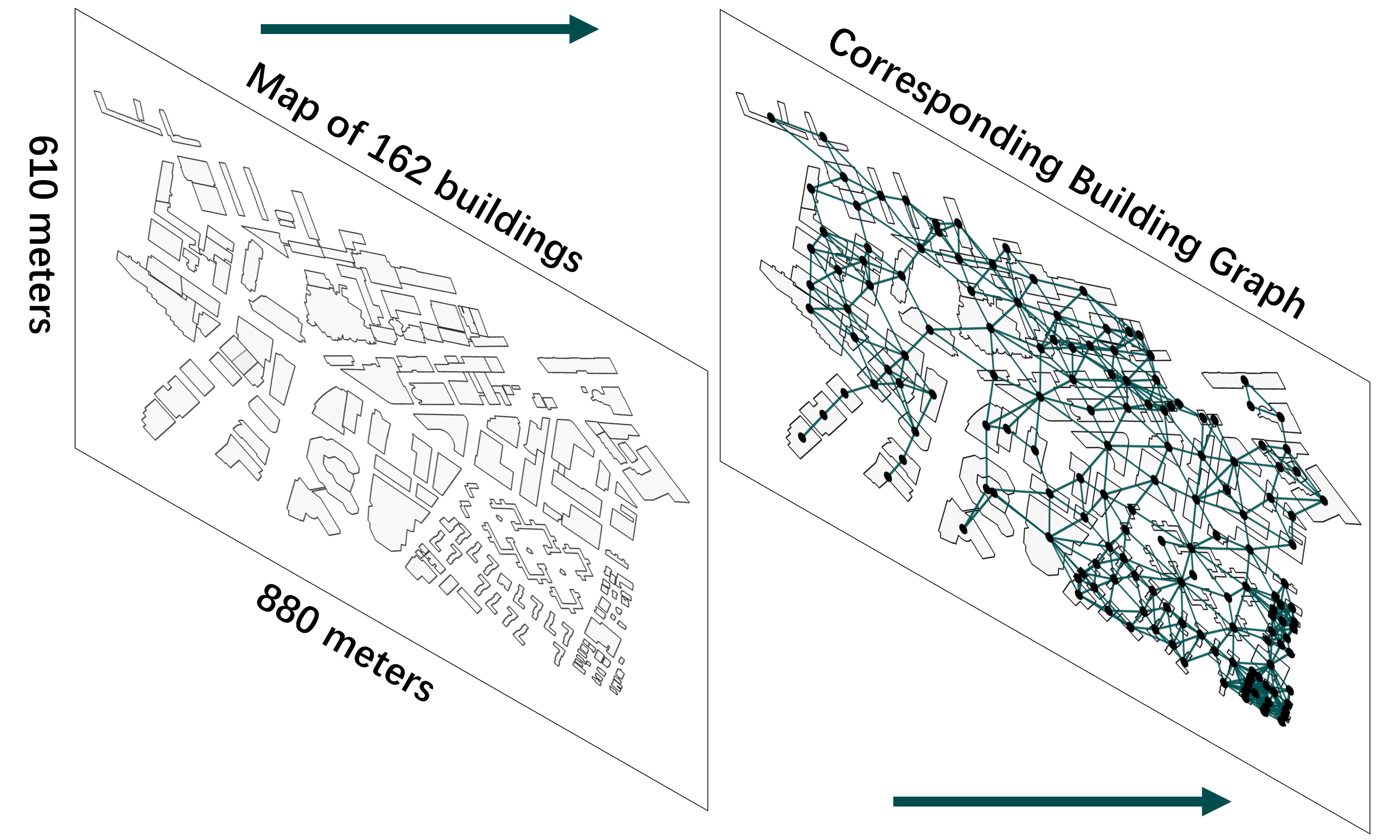}
    \caption{Converting a transformed map to Building Graph}
    \label{fig:buildingMapGraph}
\end{figure}

Given a map of buildings with known locations and footprints, we produce a building graph, $G$, whose vertices $V=\{b_1,b_2,...,b_n\}$ are buildings, each with a unique ID. Two vertices are connected by an edge if they are within a nominal Wi-Fi range, say 100 meters. In practice, not all edges are usable (more on that below), and there may be some buildings whose Wi-Fi devices can communicate with each other at greater distances, but we ignore that possibility. 


Importantly, in graph $G$ there is no knowledge of the positioning or even the presence of any actual Wi-Fi devices; it encodes only buildings. We expect with good reason that buildings will have Wi-Fi devices, particularly static Wi-Fi APs and often also mobile devices if people are present (which may not always be true, of course). To compute usable paths, we need to set reasonable edge weights, where each edge weight captures the likelihood of a packet being sent from a building (vertex) reaching the building at the other end of the edge. 

Because connectivity is a decreasing function of distance, a natural edge cost has the form $d^k$ where $d$ represents the closest distance between the two buildings and $k \geq 1$ is a hyperparameter. The larger $k$ is, the more likely that the selected path will traverse a sequence of buildings close to each other, while smaller values of $k$ prefer longer individual links. The reason is that for any three vertices $A, B, C$ with pairwise distances $d_{AB}, d_{BC}, d_{CA}$, where $d_{CA}$ is the largest of the three numbers, $d_{CA}^k - (d_{AB}^k + d_{BC}^k)$ is an increasing function of $k$; that is, as $k$ grows, the direct link from $A$ to $C$ has a higher cost than going from $A$ to $C$ via $B$. For example, when $k = 1$, due to the triangle inequality, it is always preferable to go directly; for $k=2$ it is preferable to go via $B$ if $B$ is inside the circle of diameter $AC$; and as $k$ grows, the path via $B$ becomes the lower cost path for more and more node placements. Indeed, at an extreme as $k \rightarrow \infty$, we have proved that the chosen path is always from the {\em minimum spanning tree} (MST) of the graph; one may think of this path as the {\em safest path} in the sense that it maximizes the probability of packet delivery in our model, but all traffic concentrates on the links of the MST.


We empirically tested the performance of different $k$'s for different cities, finding that $k=10$ provides a reasonable balance between reliable delivery and avoiding traffic concentration on a small fraction of available edges (building-to-building wireless links). 


\subsection{Conduits to Improve Delivery}
\label{s:conduit}

\begin{figure}[t]
    \centering
    \includegraphics[width=\linewidth]{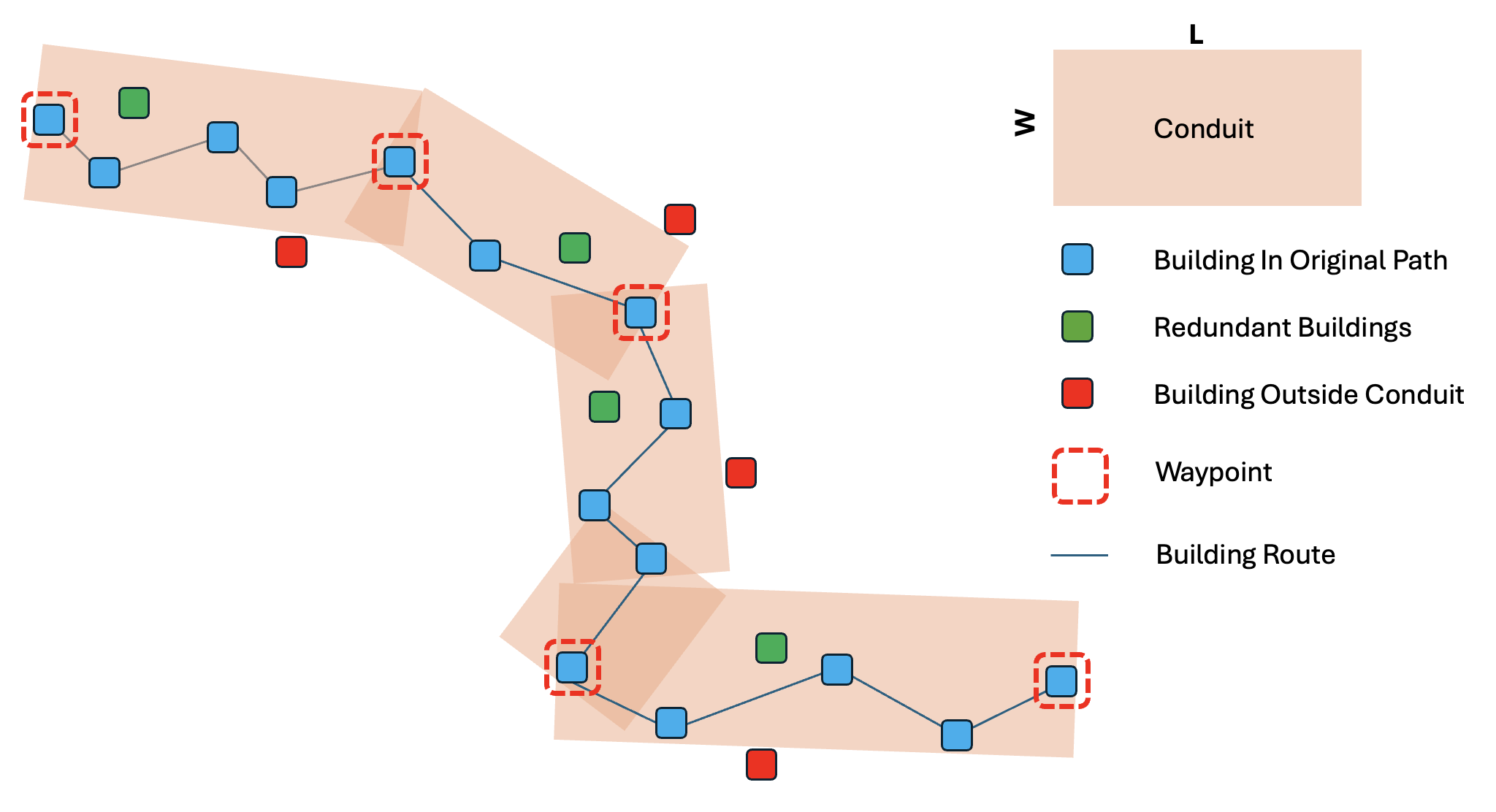}
    \caption{``Conduit'' that defines the boundary within which devices should rebroadcast a given packet. The conduit, bounded by waypoints, simplifies the route a packet should follow.}
    \label{fig:conduit-waypts}
\end{figure}

A challenge with this source routing approach is that when a source computes a path it expects that there is some working Wi-Fi link between nodes in any two successive buildings on the path. However, since devices can be randomly placed within a building (or indeed in some cases a building may have no active devices), some expected links may not be usable. To handle this issue, instead of only allowing devices within the buildings of the chosen path to forward (rebroadcast) a packet, we allow nearby buildings that are within a set distance away from the path to also forward packets. 

We implement this idea by creating a rectangle of width $W$ between two buildings $b_1$ and $b_2$, as shown in Figure~\ref{fig:conduit-waypts}. We define this \textbf{conduit} as the region containing all buildings whose centroids are within $\frac{W}{2}$ distance away from the line segment from $b_1$'s centroid to $b_2$'s centroid. We call $W$ the {\em conduit width}.

Now, once a device receives a packet, instead of checking if its building ID $b$ is in the packet's specified path, $P$, a device now checks if it's building's center point is in a conduit of $P$ by checking it's distance from the line from $P_1$ to $P_2$ is less than $\frac{W}{2}$. If, and only if it is, does the device rebroadcast the packet. This process only consumes 10 arithmetic operations.

\subsection{Waypoints to Compress Source Routes}
\label{s:waypoints}

The next issue we must resolve is that every packet must contain all the buildings on the path, $P$, which does not scale well. To reduce the size of the packet header, we introduce a compression scheme that creates a new compressed path, $P_C$, which consists only of buildings in $P$ wherever the direction of traversal changes. We call such buildings {\em waypoints}. Note that we don't lose any information about the path when we encode it using only waypoints. 

To identify the waypoints of $P$, we iteratively test if the conduit created from buildings $b_0$ and $b_i$ contains {\em all} the centroids of the buildings in $P$ from $0$ to $i$. If they are, then we continue the test by incrementing $i$. If not, then we set a waypoint at this turning point, $b_i$, and repeat the process from this new conduit origin. 
Figure~\ref{fig:conduit-waypts} shows an example.





\subsection{Grid-based Addressing and Routing}
\label{s:gridrtg}

Waypoints significantly reduce the size a packet header compared to specifying every building on the path---by a factor of $4\times$ in our tests on real cities, from 40-80 entries (each 2 bytes long) to 10-20 entries. But the header size still grows with the length of the path and the size of connected components within a city. Moreover, requiring each source to store the entire map and compute minimum-cost paths is not scalable. 

We solve this problem with a new mechanism: we replace source routing with a {\em grid-based topological addressing scheme} for buildings and construct hierarchical routing tables that devices can use to broadcast packets by performing a longest-prefix match similar to IPv4 and IPv6 destination lookups~\cite{sklower1991tree,srinivasan1999fast}. In addition, we show how a {\em central server} can create routing tables for each building and distribute these tables to each device in advance of them being necessary to be used, eliminating the need for a distributed method. Our addressing scheme compactly encodes geographic regions and buildings within a city; it is not intended to address devices or their network interfaces (nor is that necessary for the protocol).

Each routing table is intended to contain information that allows a Wi-Fi device to determine the next waypoint to use for any given packet destination address. These routing tables grow only as $O(\log S)$, where $S$ is the area of the city, independent of the number of devices in the network. Using routing tables, the information required in each packet header is the packet's destination address, the previous waypoint, and the next waypoint. We need the previous and next waypoints for a device to check if it is within the conduit, the next waypoint to also direct the packet, and the destination so the destination can receive and process the packet. It is important to note that the size of the packet header is now constant and no longer depends on the number of waypoints on the path. 

\subsubsection{Grid-based addressing} 
\label{s:gridaddr}

\begin{figure}[t]
    \centering
    \includegraphics[width=\linewidth]{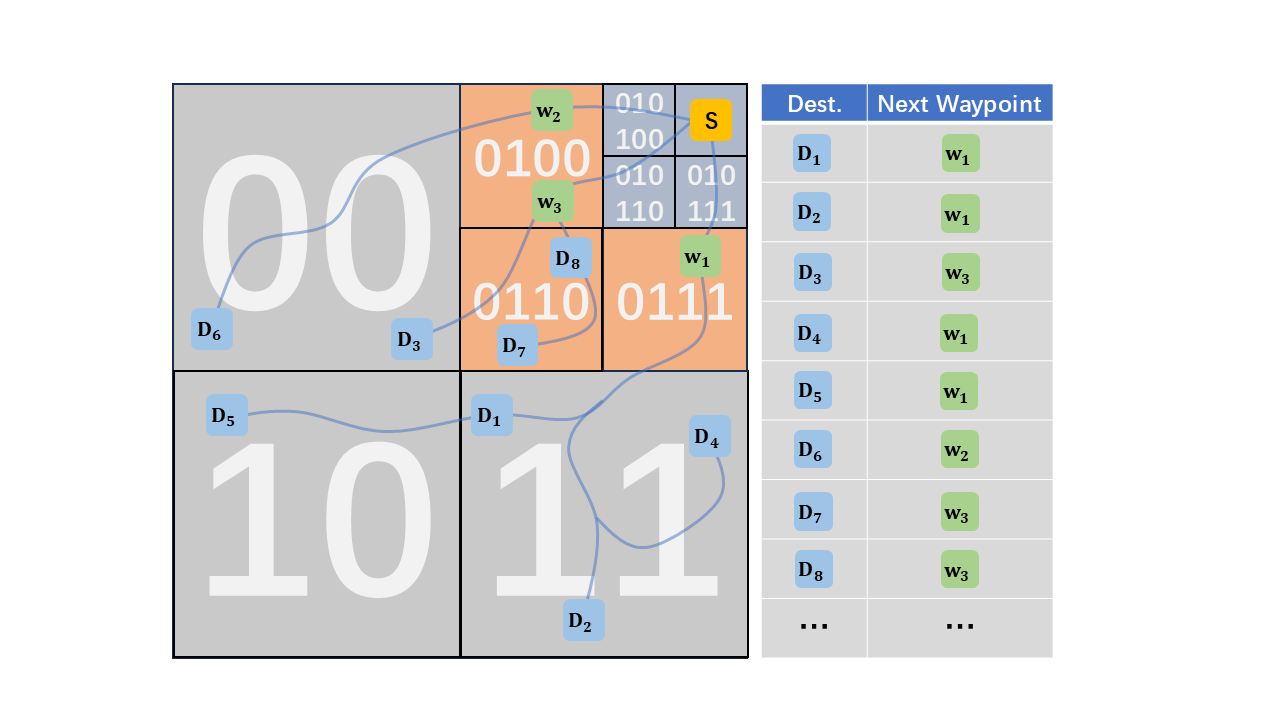}
    \caption{\textbf{Grid-based addressing and routing table.} Addresses are assigned based on buildings' location. Source \textit{S} stores all (destination building, next waypoint) pairs in the routing table.}
    \label{fig:routing-table-uncondensed}
\end{figure}

\begin{figure}[t]
    \centering
    \includegraphics[width=\linewidth]{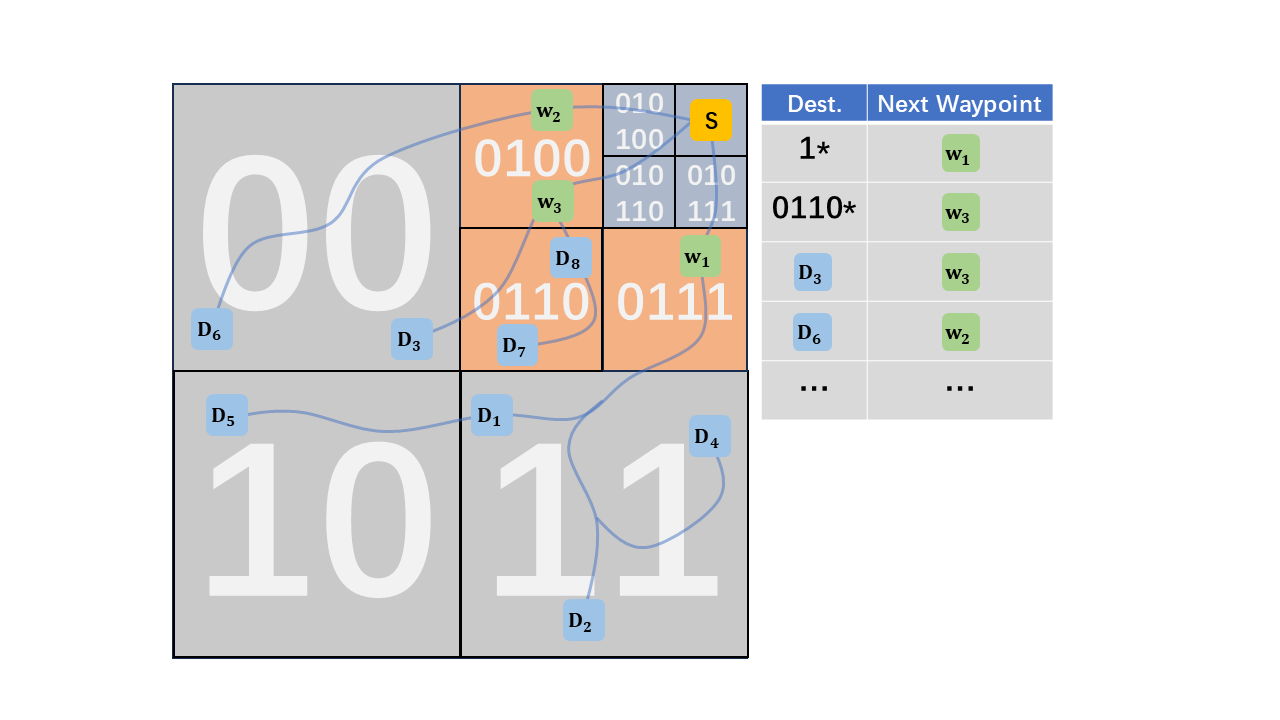}
    \caption{Condensed routing table of source \textit{S}. We can combine destination entries by specifying a coarser gridID if all destinations in that coarser grid have the same next waypoint.}
    \label{fig:routing-table-condensed}
\end{figure}

We adopt a hierarchical spatial addressing scheme where each building is given a \texttt{gridID}, which is a bitstring that compactly encodes its position on the map. For any city, we assign addresses to regions by recursively dividing the map in half along both the $x$ and $y$ axes as shown in Figure~\ref{fig:routing-table-uncondensed}. This two-dimensional recursive decomposition is similar to the Grid Location Service~\cite{li_gls}, but with the critical difference that in \name we use topological identifiers unlike in GLS where each grid is named non-topologically. This approach repeatedly creates four grids out of any given grid until the smallest grid size is reached when the length and width of the grid are roughly the nominal Wi-Fi range of $\approx$ 100 meters. 

Within this smallest grid, which we call a {\em cell}, we assign a unique ID to each building, so the last few bits of the addresses of each building in the cell are different (but they all share the same prefix). This approach ensures that every cell and every building in the mapped region has a unique address whose length is logarithmic in the size of the region. In addition, neighboring cells (and the buildings in them) share long prefixes of their grid addresses. 

\begin{figure}[t]
    \centering
    \includegraphics[width=\linewidth]{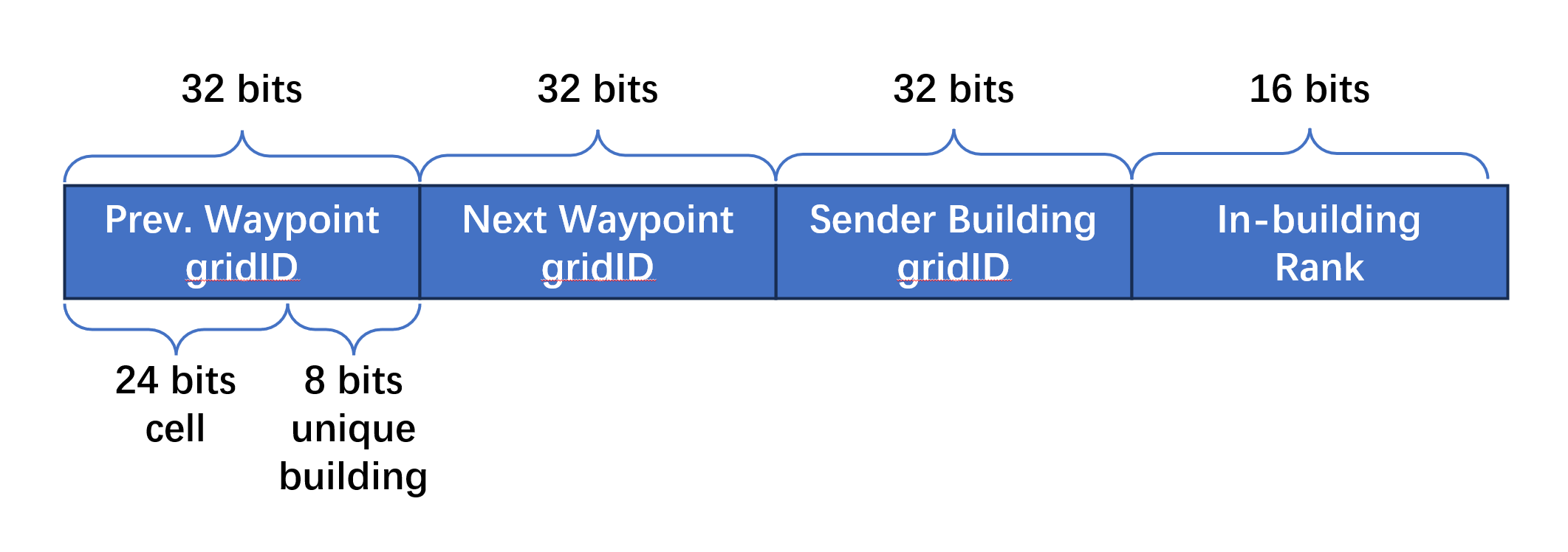}
    \caption{\name's packet header.}
    \label{fig:packet-header}
\end{figure}

\subsubsection{Route calculation}

How should \name now compute routes with this addressing scheme? Rather than have devices in buildings propagate routing messages, we note that buildings don't change all that often (unlike the mobile devices in them). This observation means that we can precompute routes at a central route server and distribute the resulting routing tables to building devices periodically. A plausible approach might be to precompute routes between every pair of buildings, determine what waypoints are used by each building to reach any given destination, and then condense the routing table using standard techniques used in IP forwarding table data structures.

However, this precomputation is impractical for every pair of buildings in a city because all-pairs-minimum-cost-path computation in an $n$-vertex graph takes $O(n^3)$ time, and a city such as New York City has a million buildings. For better scalability, instead of precomputing paths for each source-to-destination building pair, \name computes paths for every {\em cell-to-cell pair} by picking a random building in each cell; as most cells in cities contain 10-20 buildings, we find that this approach provides a 1000$\times$ speedup (denser locations have greater speedups). We also exclude each cell without a building from this computation.

The route server uses these precomputed paths to generate a routing table for each destination cell. Each entry in this table contains only the next waypoint for a given destination cell. 
At the end of this stage, buildings along the randomly selected paths will have entries to a given destination cell, and others (several) will not. 
To populate the missing entries, the route server simply copies the route entry from one of the buildings in the cell (and we know there will be at least one such building) to the other buildings within the same cell. There may be multiple buildings in the same cell with different routes to a given destination cell. In that case, we use the waypoint closest to the destination. Given that each cell is not that much larger than the typical Wi-Fi range, this optimization isn't important. Finally, we may have infrequent pathologies where a building may have a waypoint within the same cell; in this case, we ensure that the waypoint in question has a waypoint of its own that is outside the cell by running a path computation from that waypoint to the destination to force it to have an external waypoint. 
Note that while it is possible for all buildings within a cell to have the same waypoint for any given destination, in some cases they will not all share the same waypoint. This is a good feature as it provides a diversity of paths. Moreover, as explained above, at least one of the buildings will have a waypoint outside the cell. We handle the base case of a destination within the same cell by adding explicit entries as all such destinations will be within one or two Wi-Fi hops by construction.



\subsubsection{Condensing the routing table}
As described thus far, the routing tables are not compact; they will have as many entries as cells, which could be hundreds of thousands of entries in some cities and consume significant memory and computation on Wi-Fi devices.  However, we find that many entries in the table are redundant, as many different destination cells often share the same next waypoint within a cell. This is where our addressing scheme from \S\ref{s:gridrtg} is highly beneficial because we can use standard IP forwarding table compression algorithms~\cite{gabor_ip_forwarding_compress, elliot_compression, yang_compression, richard_compression} to reduce the table sizes by several orders of magnitude. Our implementation uses the method of Draves et al.~\cite{richard_compression}. In \S\ref{s:eval}, we show that when compressed, the maximum size of the routing table for a city like Rio spanning 70 km x 34 km with 115k buildings is only 639 entries, consuming only 5.1KB worth of storage.

\begin{figure*}[ht]
\begin{minipage}{\textwidth}
\begin{subfigure}{0.49\textwidth}
\includegraphics[width=\textwidth]{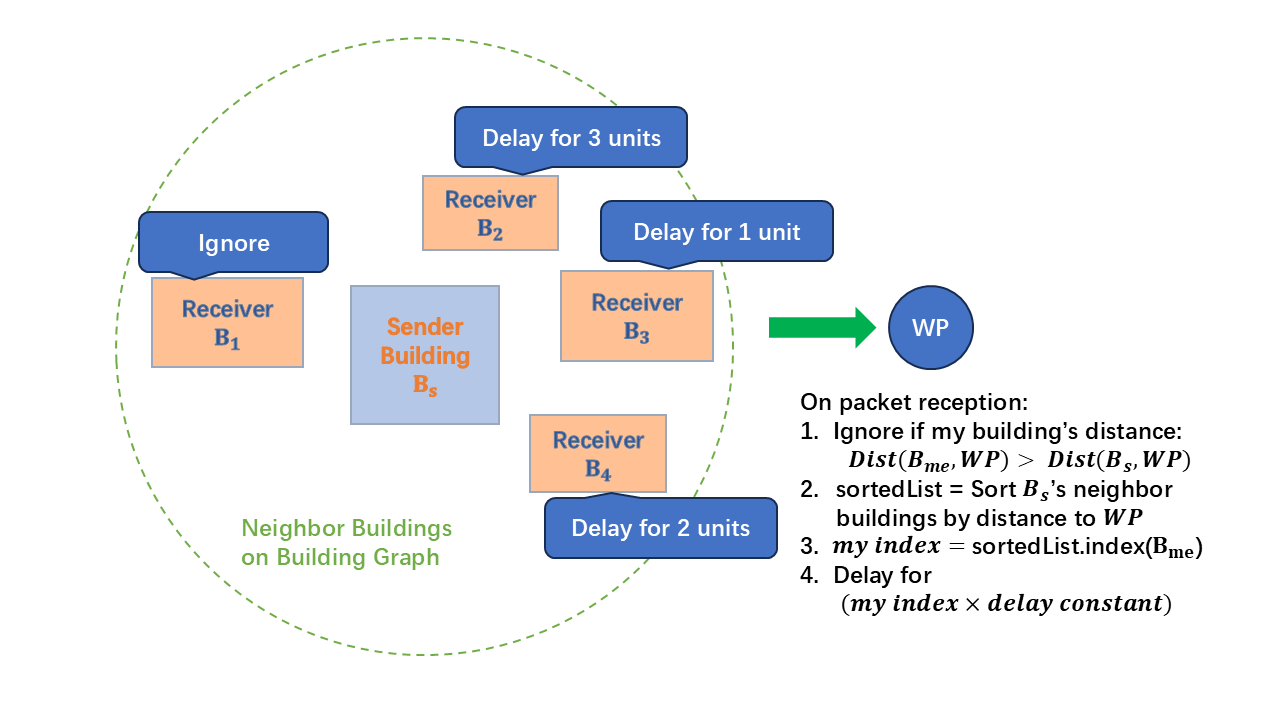}
\caption{Delay computation.}
\label{fig:1}
\end{subfigure}
\begin{subfigure}{0.49\textwidth}
    \raggedleft
\includegraphics[width=\textwidth]{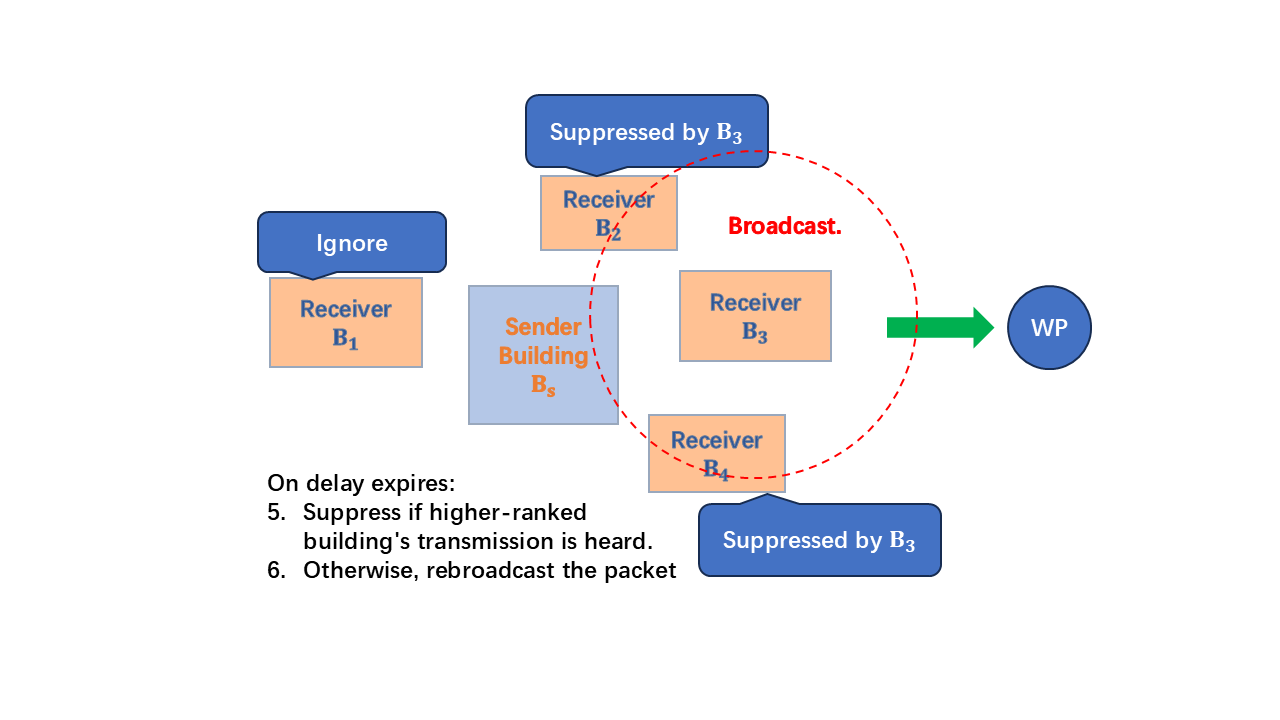}
\caption{Transmission suppression.}
\label{fig:2}
\end{subfigure}
\caption{Inter-building suppression.}\vspace{-0.1in}
\label{fig:suppressionInter}
\end{minipage}
\end{figure*}
\subsection{Suppression}
\label{s:suppression}

Allowing every device in the conduit to broadcast an overheard packet will cause huge overhead~\cite{citymesh}.
\name introduces a new suppression protocol with two distinct mechanisms:
1) \emph{inter-building suppression} to reduce the number of transmitting buildings, i.e., for determining a broadcast priority between buildings and 2) \emph{in-building suppression} to reduce broadcasts within a building.
The high-level idea is that when a device receives a packet, it calculates a delay while continuing to listen for whether a ``better'' device has broadcast the same packet. If that occurs before the timer expires, then the original device suppresses this packet broadcast; if not, then it broadcasts. Both suppression components are distributed methods with no explicit central coordinator.

\subsubsection{Inter-building suppression}


For inter-building suppression, \name's goal is to prioritize the building whose transmission (by which we mean some device in the building) is expected to make the most progress towards the destination, among multiple buildings whose devices may have heard the same packet.

When a building $S$ broadcasts a packet, each receiving device (in neighboring buildings) uses the relevant portion of the building graph\footnote{A local version of the graph containing the building's neighbors and their neighbors is sufficient for this task} to calculate a score to decide if it should rebroadcast the packet.
This score is a function of the distance between the building $S$ and the set of $S$'s neighboring buildings, $N_S$. 
Each receiver calculates a relative rank among the entries of $N_S$ by sorting each building, $b \in N_S$, by the distance to the waypoint specified in the packet. Note that because the waypoint is encoded as a grid location, this calculation is straightforward.

The neighbor closest to the next waypoint is given the highest score and delay of 1 unit, while the furthest building from the waypoint will have a score equal to the number of elements in $N_S$ and a proportionate delay. Any building further from the waypoint than $S$ will not broadcast the packet. Finally, we scale the delay for this inter-building suppression mechanism by setting the unit delay so that it is always greater than the maximum delay for in-building suppression (described below). 
As a result, the devices within the same building will resolve their suppression first before the packet starts to be suppressed at the inter-building level.

\subsubsection{In-building suppression}
\begin{figure}
    \centering
    \includegraphics[width=0.4\textwidth]{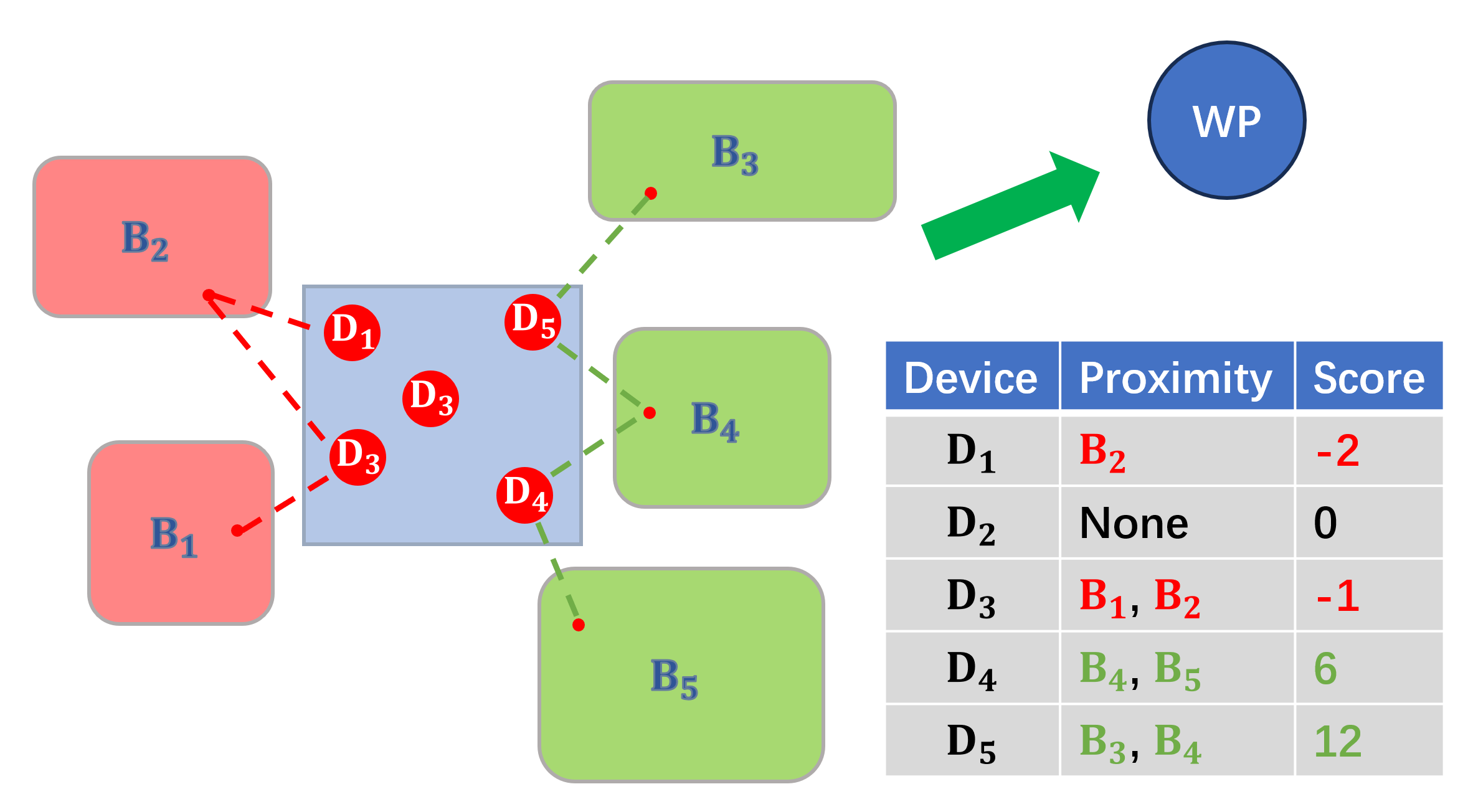}
    \caption{\small In-building suppression asks every device in the building to record the data packet sent from other buildings. It then use this record to compute the delay. Given the waypoint on the position, the red buildings are further to the waypoint than the current building, and green buildings are closer.}
    \label{fig:suppression-in-score}
\end{figure}

The core challenge with in-building suppression is that the Wi-Fi devices have no specific location information. We considered introducing such a mechanism using prior techniques, but found that they lack sufficient precision and are resource-intensive. Moreover, we show that this is not required. Each Wi-Fi device only knows which building it is within. 

By overhearing packet broadcasts from other devices, a device can instead estimate the number of distinct buildings it is able to hear in any recent interval of time without any additional beacons or control messages. If a device can see many better-positioned neighboring buildings, then within its building it should be given a higher preference to transmit. Conversely, a device that sees a large number of worse-positioned neighbors should suppress the broadcast of this packet more aggressively. In addition, the device seeing the best-positioned neighbor should be given a higher priority than other devices as this means this node can make the most progress towards a destination as possible. This approach allows for easy storage and transmission of this information without requiring any knowledge of the link quality between devices. 

To calculate an in-building rank, $R$, a device $d$ in building $b$ uses two sets: $N_b$, the set of neighboring buildings of $b$, and $N_d \subseteq N_b$, where $N_d$ is the set of neighboring buildings that $d$ can hear (reach, assuming symmetric delivery for simplicity). The device sorts the elements in $N_b$ in decreasing order of distance from the next waypoint, so the best next building to the next waypoint is at the end of the sorted list. Starting with a rank of 0, for each element in this sorted list that is also in $N_d$, we add a value of $2^i$ where $i$ is the index of the element. Then, we decrease $R$ by 1 for each neighboring building that $d$ can reach that is farther away from the next waypoint. 

Once $R$ is calculated, the device adds an in-building delay of $c * (1 - \frac{\log_2{R}}{\log_2{(best\_score)}})$ (we handle any negative $R$ in a suitable way, the details aren't important here). With this method, the best node will have a delay of 0, while the worst node will have an in-building delay of $2c$.

If two devices have the same delay, we use a randomized jitter to break ties; a device that overhears another with the same rank broadcasting the packet will suppress its own broadcast.

\begin{figure*}[t]
    \centering
    \begin{subfigure}{0.32\textwidth}
        \includegraphics[width=\textwidth]{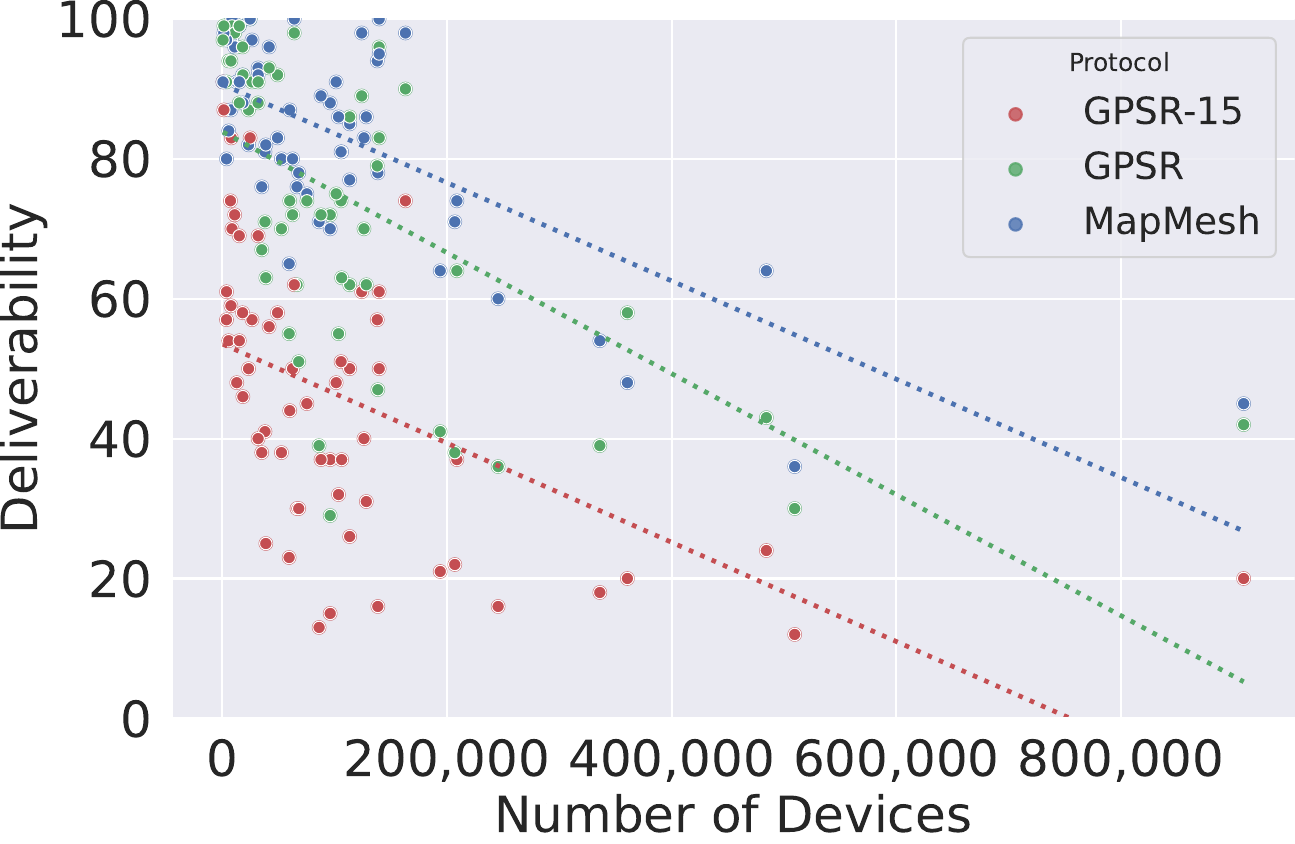}
        \caption{$\ell =$ 0.4.}
        \label{fig:2}
    \end{subfigure}
    \begin{subfigure}{0.32\textwidth}
        \includegraphics[width=\textwidth]{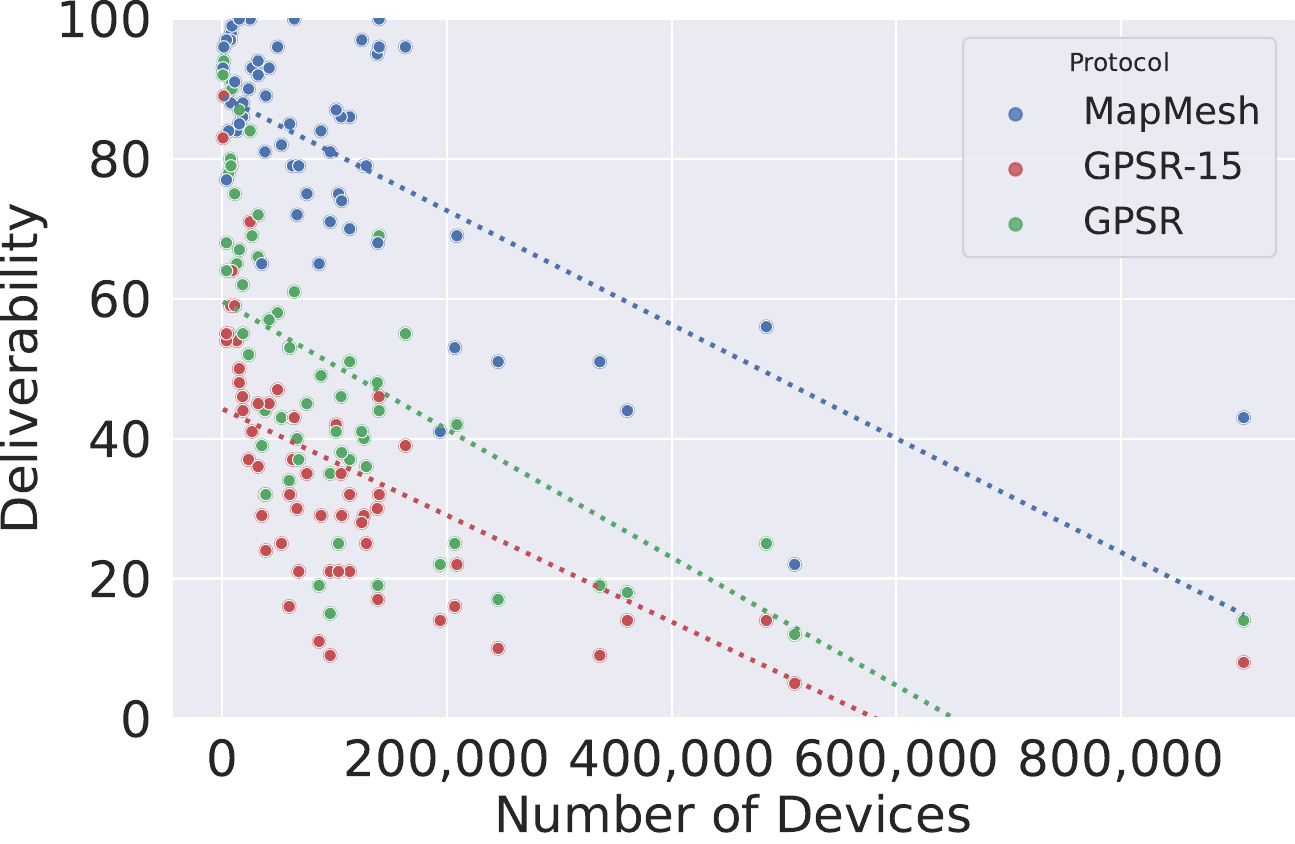}
        \caption{$\ell =$ 0.6.}
        \label{fig:3}
    \end{subfigure}
    \begin{subfigure}{0.32\textwidth}
        \includegraphics[width=\textwidth]{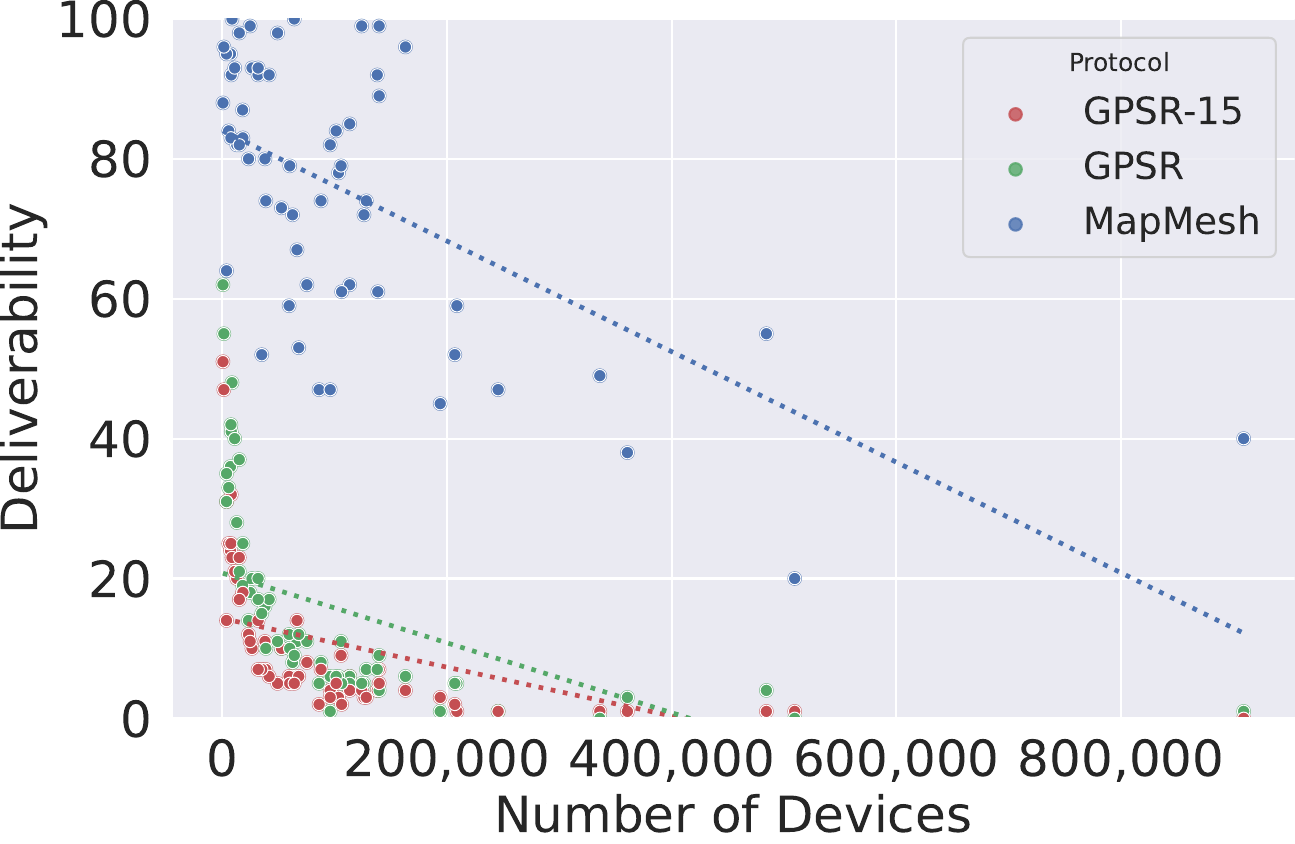}
        \caption{$\ell =$ 0.8.}
        \label{fig:4}
    \end{subfigure}
    \caption{Packet delivery rates in 60 simulated city regions for \name, GPSR without location error, and GPSR with a random uniform location error from -15 m to 15 m. Packet loss rates are uniformly picked at random between 0 and $\ell$, where $\ell$ is 0.4 (left), 0.6 (middle), and 0.8 (right). The overall conclusion is that \name significantly outperforms GPSR.
    }
    \label{fig:delivery}
\end{figure*}

\begin{figure*}[t]
    \centering
    \begin{subfigure}{0.32\textwidth}
        \includegraphics[width=\textwidth]{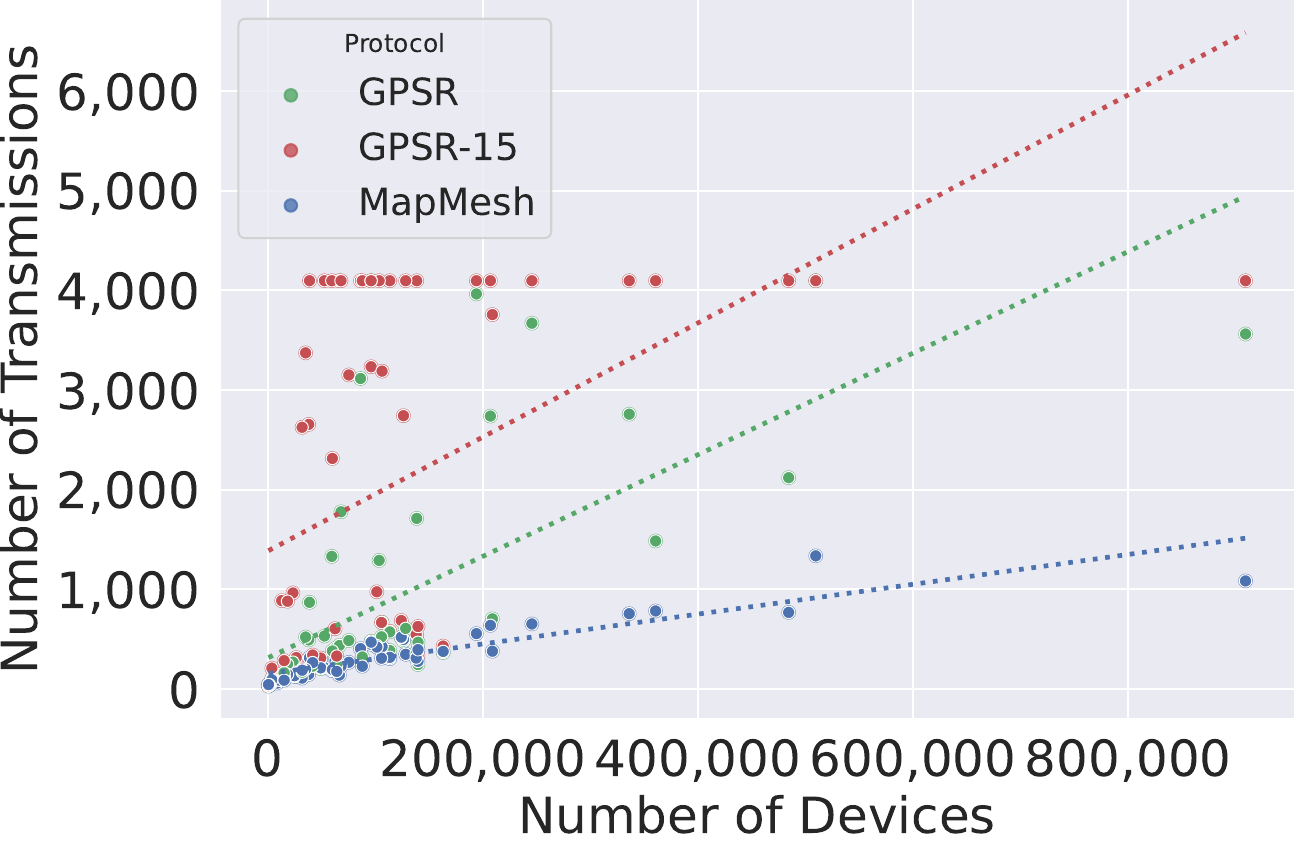}
        \caption{$\ell = 0.4$}
        \label{fig:2}
    \end{subfigure}
    \begin{subfigure}{0.32\textwidth}
        \includegraphics[width=\textwidth]{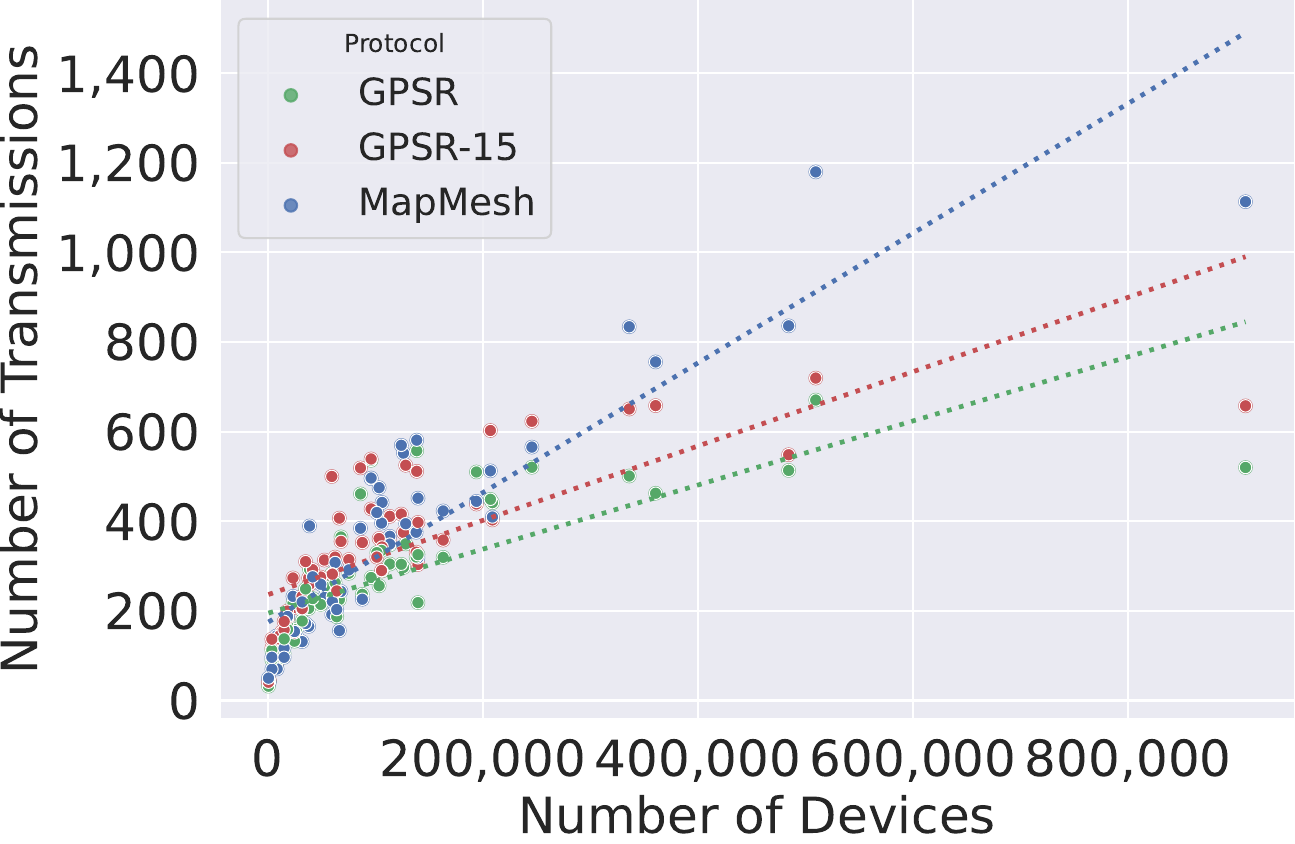}
        \caption{$\ell = 0.6$}
        \label{fig:3}
    \end{subfigure}
    \begin{subfigure}{0.32\textwidth}
        \includegraphics[width=\textwidth]{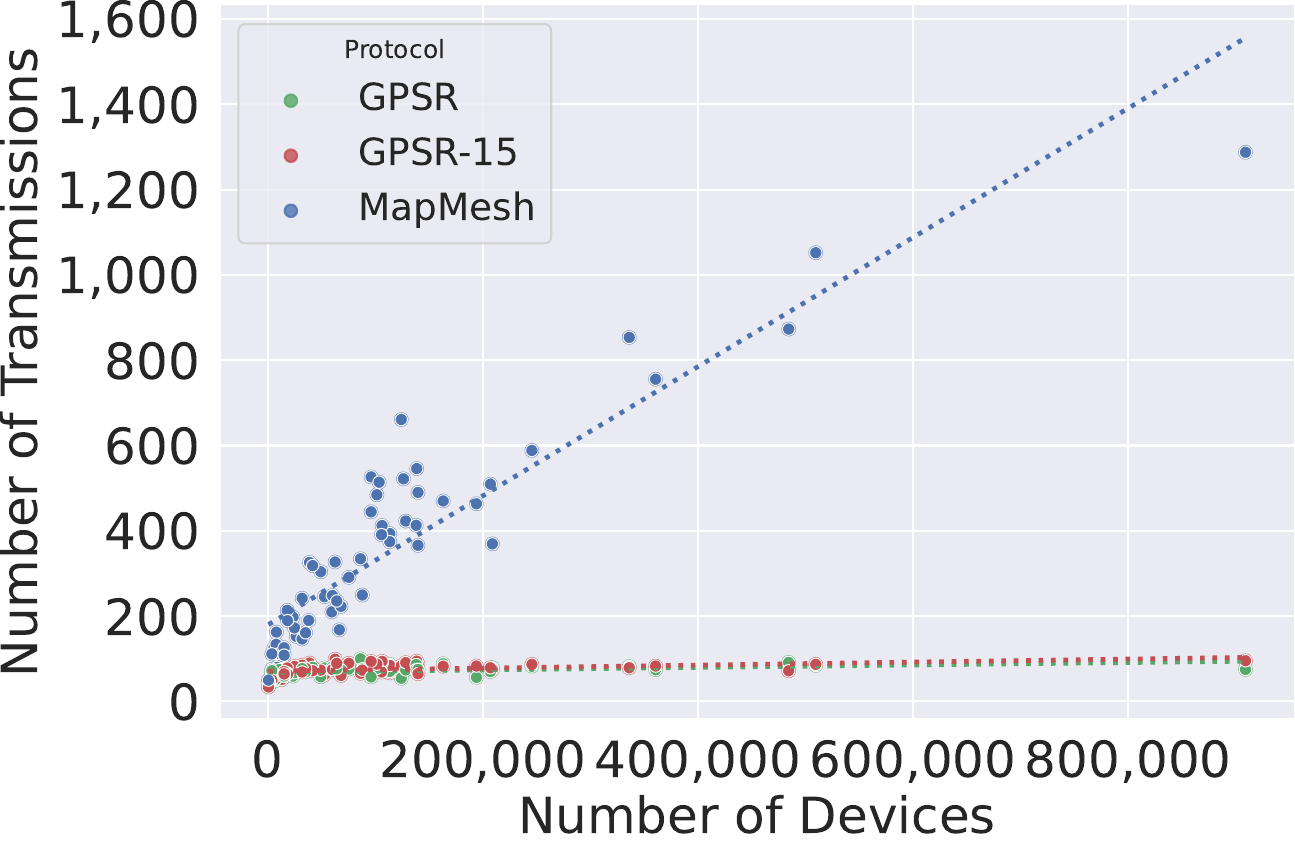}
        \caption{$\ell = 0.8$}
        \label{fig:4}
    \end{subfigure}
    \caption{Number of transmissions with a stochastic drop rate from 0 to $\ell$ equal to 0.4 (left), 0.6 (middle), and 0.8 (right). \name outperforms GPSR, often by some orders of magnitude. 
    }
    \label{fig:transmissions}
\end{figure*}

\begin{figure*}
    \centering
    \makebox[\textwidth][c]{%
        \begin{minipage}{0.32\textwidth}
            \centering
            \includegraphics[width=\linewidth]{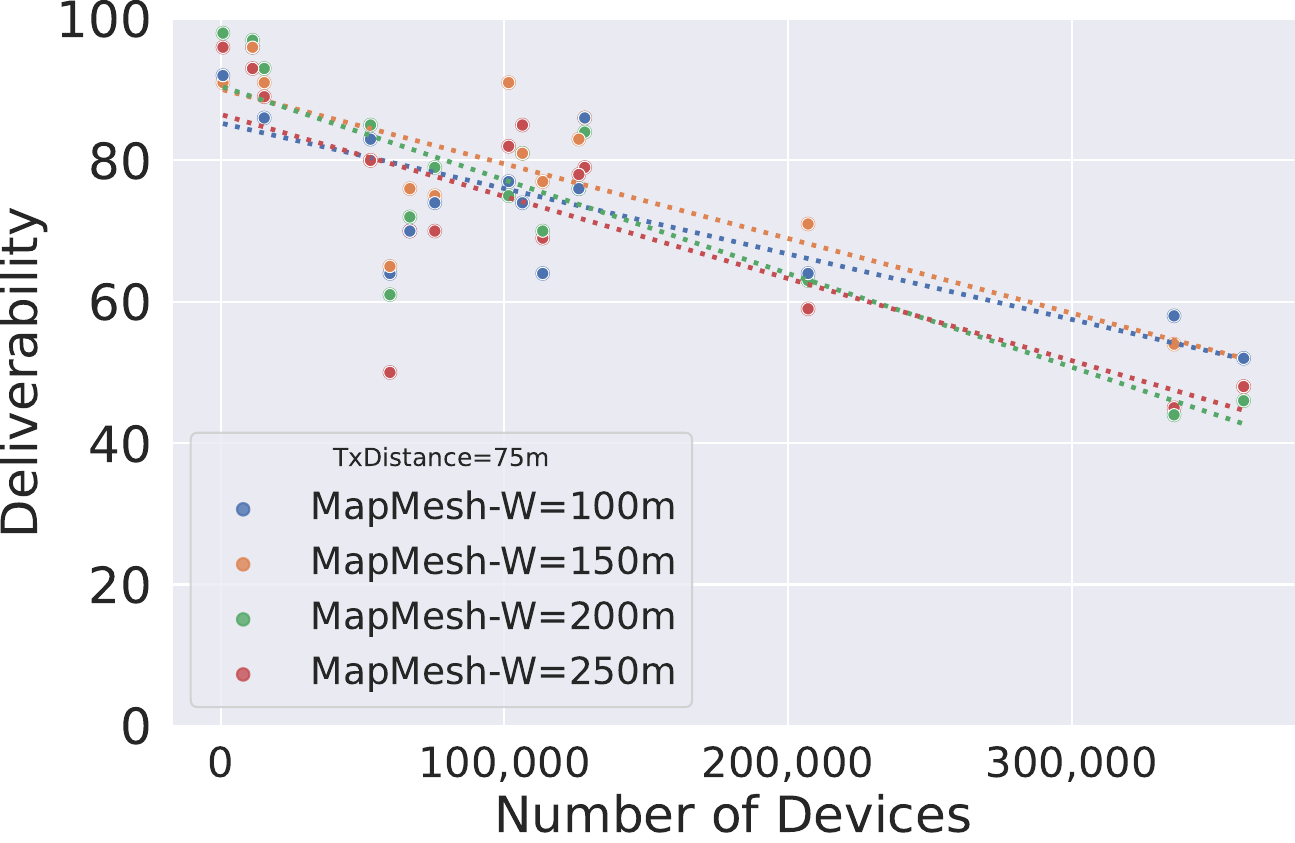}
            \caption{Variation of conduit widths. A conduit width provides additional redundancy until a point. W=150m provides the best deliverability.}
            \label{fig:conduitDelivery}
        \end{minipage}
        \hfill
        \begin{minipage}{0.32\textwidth}
            \centering
            \includegraphics[width=\linewidth]{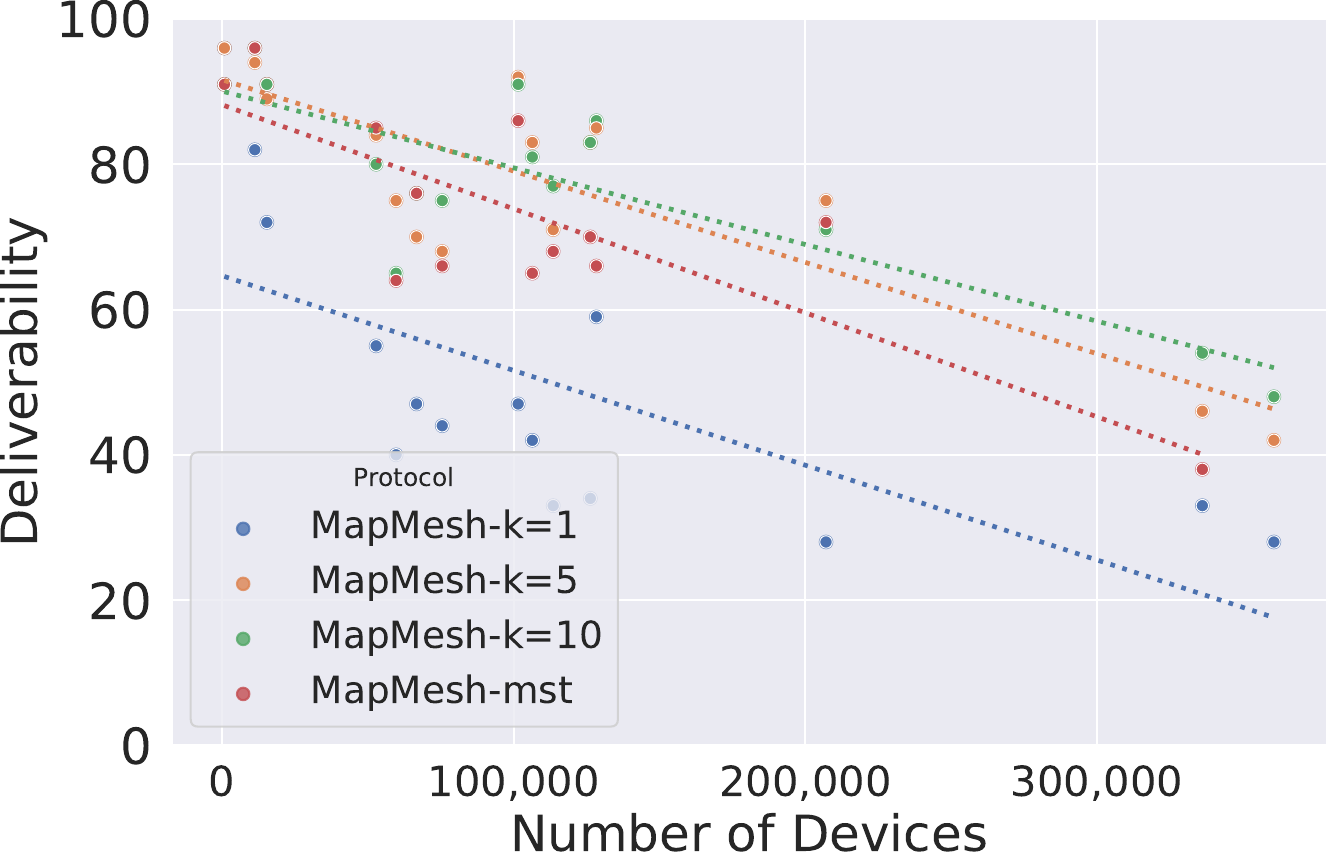}
            \caption{Variation of different k-values for path selection in 15 different cities. k=10 provides the best overall performance for deliverability.}
            \label{fig:varyingKs}
        \end{minipage}
        \hfill
        \begin{minipage}{0.32\textwidth}
            \centering
            \includegraphics[width=\linewidth]{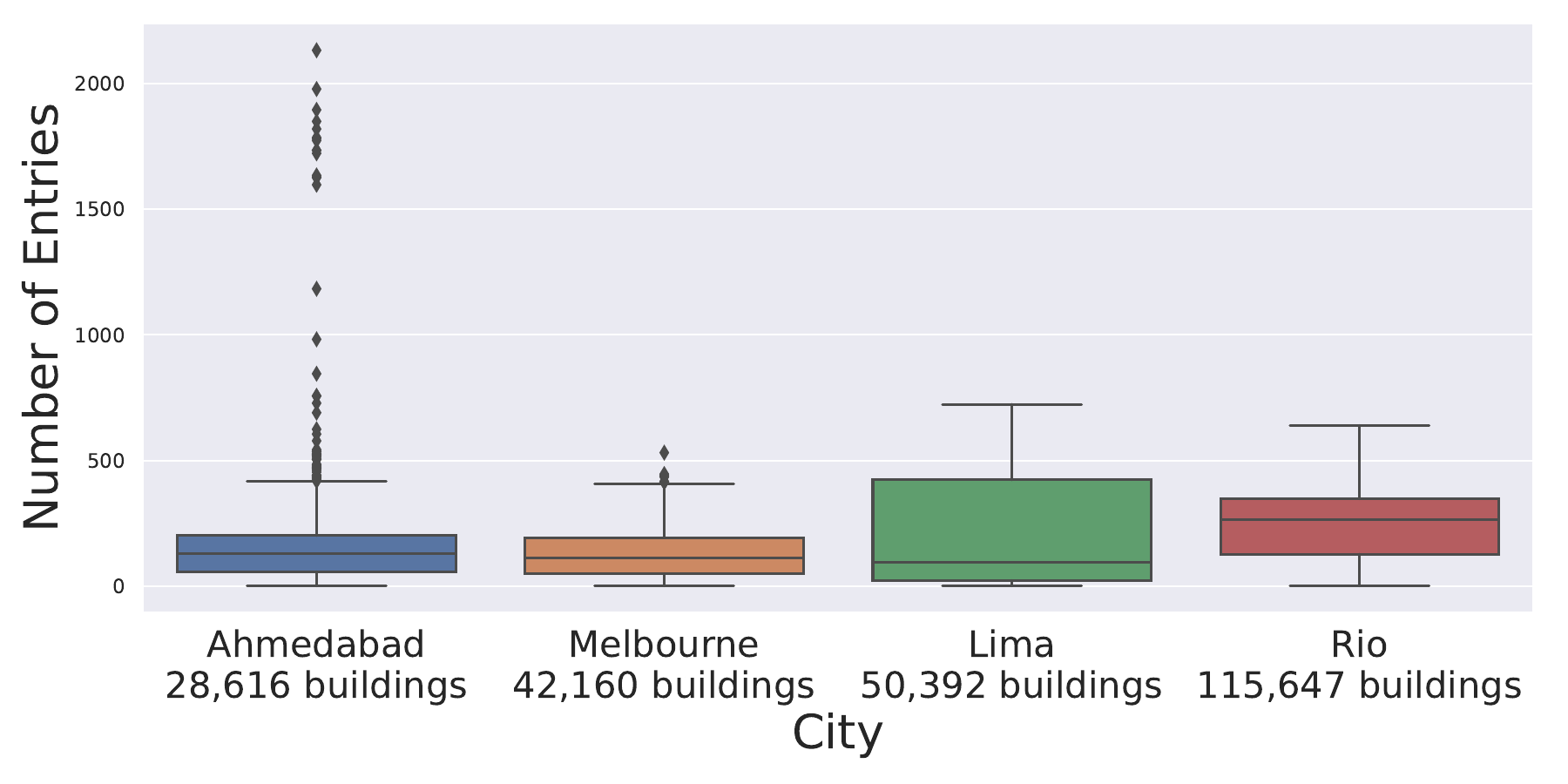}
            \caption{The routing tables can be efficiently compressed, requiring each device to store a maximum of 2132 entries - or 16 KB in Ahmedabad}
            \label{fig:tableSize}
        \end{minipage}
    }
\end{figure*}

\section{City-scale Simulation Results}
\label{s:large-scale}
\label{s:eval}

Our city-scale simulation of \name addresses the following questions:
\begin{itemize}
    \item What is the packet delivery rate of \name under varying wireless conditions in various cities?
    \item How many total transmissions does \name have, i.e., what is the bandwidth overhead?
    \item How does varying the conduit width affect performance?
    \item How does the exponent $k$ affect performance?
    \item How much routing information are devices required to store in a real system?
\end{itemize}

We use ns-3 to model a realistic deployment of wireless devices at the city scale. We report representative results from regions of 60 cities around the world. For each city, we collect the building map from OpenStreetMap (OSM)~\cite{OpenStreetMap}. We place wireless devices within each building at a density of roughly 200 m$^2$; we conducted small-scale measurements to determine that typical Wi-Fi densities in cities are usually higher than this number.

\textbf{Packet loss model.} We used ns-3's built-in distance-based model for a typical 2.4 GHz Wi-Fi setting, which has the property that the packet loss rate is largely zero until a range of about 70 meters, and increases rather sharply to 1 within 10 meters after that. This model does not handle other stochastic losses that we see in practice including in our testbed and in prior works like Roofnet, which see loss rates across the range from 0 to 1. Thus we add atop the distance-based ns-3 model for each link a uni-directional link-layer loss rate picked uniformly at random for each packet between 0 and a maximum of $\ell$. We show results for multiple values of $\ell$ between 0.4 and 0.8. 

To validate both our algorithm implementation and underlying assumptions about Wi-Fi connectivity, we built a hardware testbed using Hak5 WiFi Pineapple Mark VII devices~\cite{hak5}. The Pineapple has three 2.4~GHz Wi-Fi radios and a single-core MIPS processor. Being based on commodity hardware, similar to that which one would find in any consumer home Wi-Fi/router, it is a suitable choice for building out a testbed for \name.
Further, as the Pineapple runs modified OpenWRT~\cite{openwrt}, it is flexible and compatible with upstream OpenWRT.

To estimate packet loss rates between buildings, we deployed eight Pineapples devices across three buildings on a university campus.
We placed the devices across different floors in nearby buildings and configured one of their wireless network interfaces as a mesh point.
Each device broadcasted a single UDP packet with a unique identifier over a dedicated 802.11 mesh point interface every 5 seconds and logged these transmissions.
Simultaneously, all Pineapples listened on the same interface for broadcasted packets and logged the unique values and timestamps of received packets. The results we obtained justify our choice of experimental parameters.

\textbf{Packet deliverability.} In Figure \ref{fig:delivery}, we pick 100 source-destination pairs at random and send a packet for each pair. We compute the packet delivery rate (fraction of packets delivered) and the overhead (total number of broadcasts required). 
While \name uses link-layer broadcasts and cannot benefit from link-layer retransmissions, the GPSR variants can, up to 8 retransmits per packet. For GPSR variants, we set a TTL of 4096. 

Figure~\ref{fig:delivery} compares the packet delivery rates for 60 cities. We compare different packet loss rate scenarios across \name,  GPSR without location error, and GPSR with a random uniform location error from -15 m to 15 m (GPSR-15) for packet loss rates uniformly picked at random between 0 and $\ell$, where $\ell$ is 0.4 (left), 0.6 (middle), and 0.8 (right). 
When packet loss rates are lower, \name slightly outperforms GPSR with accurate location information but significantly outperforms it as packet loss rates rise. 


\textbf{Number of transmissions.} Figure~\ref{fig:transmissions} compares the number of retransmissions across different routing schemes for 60 cities. As shown in Figure~\ref{fig:transmissions}(a), when the link layer loss rate ($\ell$) is 0.4, \name outperforms GPSR by a factor of 3$\times$ in our biggest city, with a maximum factor of 7$\times$. This is due to GPSR's lack of global knowledge making it to become trapped in its recovery mode. Similarly, we observe that \name induces up to 28$\times$ fewer transmissions compared to GPSR with 15-meter location error. When the link layer loss rate is increased to 0.8 (Figure~\ref{fig:transmissions}(c)), GPSR has lower transmissions compared to \name simply because it fails to deliver more than 80\% of packets.

\textbf{Conduit width.} We repeated the simulations in 15 cities varying the conduit width, as shown in Figure~\ref{fig:conduitDelivery}. We observe that increasing the conduit width provides an additional redundancy. However, as the conduit width increases, the original computed path is less likely to be traversed, and more drops will occur. For all other plots, we select a conduit width of 150m.

\textbf{K value.} In \S\ref{s:rtgmetric}, we used the exponent $k$ to control how paths are selected. An increasing value of $k$ will select buildings that are closer together. To determine a proper value of $k$, we repeated the simulation over 15 cities with varying values of $k$. Figure~\ref{fig:varyingKs} shows the deliverability rate for differing $k$'s. We observe that $k=10$ provides the best deliverability, and outperforms $k=1$ and the MST ($k\rightarrow\infty$). This is because $k=1$ selects a path that encourages the packet to hop over distant buildings while the MST selects a path that is too long, accumulating the chance of a dropped packet over the long path. For all other results, we select $k=10$.

\textbf{Routing table size.} Figure~\ref{fig:tableSize} plots the size of 1000 randomly selected buildings in four representative cities, Ahmedabad, Melbourne, Lima, and Rio. Each entry in the table consumes up to 8 bytes. For Ahmedabad, the mean number of entries is 179 (1432 bytes) in a network of nearly 100k buildings. The maximum size is 2132 entries (16 KB). For Melbourne, Lima, and Rio the mean number of entries are 130 (1040 bytes), 245 (1960 bytes), and 208 (1664 bytes), respectively. These findings demonstrate \name's compact routing tables even for sizable networks.


\section{Conclusion}

  We presented \name, a city-scale scalable wireless routing system suited for disaster recovery and emergencies. When wide-area connectivity is unavailable or significantly degraded, \name enables static access points and mobile devices equipped with Wi-Fi in a city to route packets via each other for intra-city connectivity. The chief contribution of our work is a new routing protocol that scales to hundreds of thousands of nodes, a significant improvement over prior work on wireless mesh and mobile ad hoc networks. Our approach uses information about the placement and geometry of buildings from widely available urban-area maps to compute paths in a scalable way. An example simulation result shows sufficient packet delivery rates at modest packet overhead even with high link-layer packet loss rates with only a few hundred routing table entries for a network with over 900k devices. 
  



\label{bodypage}

\begin{acks}
This work was supported by DARPA Contract HR001120C0191 and Sloan fellowship FG-2022-18504.
\end{acks}

\clearpage

\bibliographystyle{abbrv}
\bibliography{reference}

\begin{thebibliography}{10}

\bibitem{afanasyev2008analysis}
M.~Afanasyev, T.~Chen, G.~M. Voelker, and A.~C. Snoeren.
\newblock Analysis of a mixed-use urban wifi network: When metropolitan becomes neapolitan.
\newblock In {\em ACM IMC}, 2008.

\bibitem{odlw}
A.~N. Al-Khwildi, S.~Khan, K.~Loo, and H.~S. Al-Raweshidy.
\newblock {On-Demand Link Weight Routing Protocol with Cross-Layer Communication}.
\newblock In {\em 18th Intl. Symp. on Personal, Indoor and Mobile Radio Comm.}, 2007.

\bibitem{qolsr}
H.~Badis, I.~Gawedzki, and K.~Al~Agha.
\newblock {QoS routing in ad hoc networks using QOLSR with no need of explicit reservation}.
\newblock In {\em 60th IEEE Vehicular Technology Conf.}, 2004.

\bibitem{dream}
S.~Basagni, I.~Chlamtac, V.~R. Syrotiuk, and B.~A. Woodward.
\newblock {A distance routing effect algorithm for mobility (DREAM)}.
\newblock In {\em MobiCom}, 1998.

\bibitem{batman_adv}
{B.A.T.M.A.N. Advanced Development Team}.
\newblock {B.A.T.M.A.N. Advanced (batman-adv)}.
\newblock \url{https://www.open-mesh.org/projects/batman-adv}.
\newblock Accessed: 2025-01-06.

\bibitem{roofnet}
J.~Bicket, D.~Aguayo, S.~Biswas, and R.~Morris.
\newblock Architecture and evaluation of an unplanned 802.11b mesh network.
\newblock In {\em MobiCom}, 2005.

\bibitem{face_routing}
P.~Bose, P.~Morin, I.~Stojmenovi{\'c}, and J.~Urrutia.
\newblock Routing with guaranteed delivery in ad hoc wireless networks.
\newblock In {\em ACM Discrete algorithms and methods for mobile computing and communications (DIALM)}, 1999.

\bibitem{brik_measurement_2008}
V.~Brik, S.~Rayanchu, S.~Saha, S.~Sen, V.~Shrivastava, and S.~Banerjee.
\newblock A measurement study of a commercial-grade urban wifi mesh.
\newblock In {\em ACM IMC}, 2008.

\bibitem{corson_1995}
M.~S. Corson and A.~Ephremides.
\newblock A distributed routing algorithm for mobile wireless networks.
\newblock {\em Wireless Network}, 1:61–81, 1995.

\bibitem{richard_compression}
R.~Draves, C.~King, S.~Venkatachary, and B.~Zill.
\newblock {Constructing optimal IP routing tables}.
\newblock In {\em INFOCOM}, 1999.

\bibitem{msr_mesh}
R.~Draves, J.~Padhye, and B.~Zill.
\newblock Comparison of routing metrics for static multi-hop wireless networks.
\newblock In {\em Proceedings of the 2004 Conference on Applications, Technologies, Architectures, and Protocols for Computer Communications}, 2004.

\bibitem{alarm}
K.~El~Defrawy and G.~Tsudik.
\newblock Alarm: Anonymous location-aided routing in suspicious manets.
\newblock {\em IEEE Transactions on Mobile Computing}, 10(9):1345--1358, 2011.

\bibitem{freifunk}
{Freifunk Community}.
\newblock {Freifunk Mesh Networking Project}, 2025.
\newblock Accessed: 2025-01-30.

\bibitem{fusler_contention-based_nodate}
H.~Füßler, H.~Hartenstein, M.~Mauve, W.~Effelsberg, and J.~Widmer.
\newblock Contention-based forwarding for street scenarios.
\newblock {\em International Workshop in Intelligent Transportation (WIT)}, 2004.

\bibitem{star}
J.~J. Garcia-Luna-Aceves and M.~Spohn.
\newblock Source-tree routing in wireless networks.
\newblock In {\em Proceedings of the Seventh Annual International Conference on Network Protocols}, 1999.

\bibitem{lanmar}
P.~Guangyu, M.~Geria, and X.~Hong.
\newblock Lanmar: landmark routing for large scale wireless ad hoc networks with group mobility.
\newblock In {\em 2000 First Annual Workshop on Mobile and Ad Hoc Networking and Computing. MobiHOC (Cat. No.00EX444)}, 2000.

\bibitem{ara}
M.~Gunes, U.~Sorges, and I.~Bouazizi.
\newblock Ara-the ant-colony based routing algorithm for manets.
\newblock In {\em Proceedings. International Conference on Parallel Processing Workshop}, 2002.

\bibitem{ietf-manet-zone-zrp-04}
Z.~Haas.
\newblock The zone routing protocol (zrp) for ad hoc networks, 1998.

\bibitem{hak5}
Hak5.
\newblock Wifi pineapple mk. vii.
\newblock \url{https://shop.hak5.org/products/wifi-pineapple}.
\newblock Accessed: 2024-01-26.

\bibitem{OpenStreetMap}
M.~Haklay and P.~Weber.
\newblock Openstreetmap: User-generated street maps.
\newblock {\em IEEE Pervasive Computing}, 7(4):12--18, 2008.

\bibitem{rip}
C.~Hedrick.
\newblock {Routing Information Protocol}.
\newblock RFC 1058, Internet Engineering Task Force, June 1988.

\bibitem{iwata_scalable_1999}
A.~Iwata, C.-C. Chiang, G.~Pei, M.~Gerla, and T.-W. Chen.
\newblock Scalable routing strategies for ad hoc wireless networks.
\newblock {\em IEEE Journal on Selected Areas in Communications}, 17(8):1369--1379, 1999.

\bibitem{olsr}
P.~Jacquet, P.~Muhlethaler, T.~Clausen, A.~Laouiti, A.~Qayyum, and L.~Viennot.
\newblock Optimized link state routing protocol for ad hoc networks.
\newblock In {\em Proceedings. IEEE International Multi Topic Conference, 2001. IEEE INMIC 2001. Technology for the 21st Century.}, pages 62--68, 2001.

\bibitem{rahul_geographical}
R.~Jain, A.~Puri, and R.~Sengupta.
\newblock Geographical routing using partial information for wireless ad hoc networks.
\newblock {\em IEEE Personal Communications}, 8(1):48--57, 2001.

\bibitem{jerbi_improved_2007}
M.~Jerbi, S.-M. Senouci, R.~Meraihi, and Y.~Ghamri-Doudane.
\newblock An {Improved} {Vehicular} {Ad} {Hoc} {Routing} {Protocol} for {City} {Environments}.
\newblock In {\em IEEE ICC}, 2007.

\bibitem{jerbi_towards_2009}
M.~Jerbi, S.-M. Senouci, T.~Rasheed, and Y.~Ghamri-Doudane.
\newblock Towards efficient geographic routing in urban vehicular networks.
\newblock {\em IEEE Transactions on Vehicular Technology}, 58(9):5048--5059, 2009.

\bibitem{zhls}
M.~Joa-Ng and I.-T. Lu.
\newblock A peer-to-peer zone-based two-level link state routing for mobile ad hoc networks.
\newblock {\em IEEE Journal on Selected Areas in Communications}, 1999.

\bibitem{dsr}
D.~B. Johnson, D.~A. Maltz, J.~Broch, et~al.
\newblock Dsr: The dynamic source routing protocol for multi-hop wireless ad hoc networks.
\newblock {\em {Ad Hoc Networks}}, 5(1):139--172, 2001.

\bibitem{gpsr}
B.~Karp and H.-T. Kung.
\newblock {GPSR: Greedy perimeter stateless routing for wireless networks}.
\newblock In {\em ACM MobiCom}, 2000.

\bibitem{elliot_compression}
E.~Karpilovsky, M.~Caesar, J.~Rexford, A.~Shaikh, and J.~van~der Merwe.
\newblock Practical network-wide compression of ip routing tables.
\newblock {\em IEEE Transactions on Network and Service Management}, 9(4):446--458, 2012.

\bibitem{kim_geographic_nodate}
Y.-J. Kim, R.~Govindan, B.~Karp, and S.~Shenker.
\newblock Geographic routing made practical.
\newblock In {\em USENIX NSDI}, 2005.

\bibitem{kiran_experimental_2018}
K.~Kiran, N.~P. Kaushik, S.~Sharath, P.~D. Shenoy, K.~R. Venugopal, and V.~T. Prabhu.
\newblock Experimental {Evaluation} of {BATMAN} and {BATMAN}-{Adv} {Routing} {Protocols} in a {Mobile} {Testbed}.
\newblock In {\em {IEEE} {Region} 10 {Conference}}, 2018.

\bibitem{lar}
Y.~Ko and N.~H. Vaidya.
\newblock Location‐aided routing (lar) in mobile ad hoc networks.
\newblock {\em Wireless Network}, 6(4):307–321, 2000.

\bibitem{goafr_plus}
F.~Kuhn, R.~Wattenhofer, Y.~Zhang, and A.~Zollinger.
\newblock Geometric ad-hoc routing: Of theory and practice.
\newblock In {\em ACM PODC}, 2003.

\bibitem{goafr}
F.~Kuhn, R.~Wattenhofer, and A.~Zollinger.
\newblock Worst-case optimal and average-case efficient geometric ad-hoc routing.
\newblock In {\em ACM MobiHoc}, 2003.

\bibitem{liangli_gpsr}
L.~Lai, Q.~Wang, and Q.~Wang.
\newblock Research on one kind of improved gpsr algorithm.
\newblock In {\em 2012 International Conference on Computer Science and Electronics Engineering}, 2012.

\bibitem{lee_enhanced_2007}
K.~C. Lee, J.~Haerri, U.~Lee, and M.~Gerla.
\newblock Enhanced {Perimeter} {Routing} for {Geographic} {Forwarding} {Protocols} in {Urban} {Vehicular} {Scenarios}.
\newblock In {\em IEEE Globecom Workshops}, 2007.

\bibitem{gdstr}
B.~Leong, B.~Liskov, and R.~T. Morris.
\newblock Geographic routing without planarization.
\newblock In {\em NSDI}, 2006.

\bibitem{pvex}
B.~Leong, S.~Mitra, and B.~Liskov.
\newblock Path vector face routing: Geographic routing with local face information.
\newblock In {\em IEEE International Conference on Network Protocols (ICNP)}, 2005.

\bibitem{li_gls}
J.~Li, J.~Jannotti, D.~S.~J. De~Couto, D.~R. Karger, and R.~Morris.
\newblock A scalable location service for geographic ad hoc routing.
\newblock In {\em MobiCom}, 2000.

\bibitem{lochert_routing_2003}
C.~Lochert, H.~Hartenstein, J.~Tian, H.~F{\"u}{\ss}ler, D.~Hermann, and M.~Mauve.
\newblock A routing strategy for vehicular ad hoc networks in city environments.
\newblock In {\em IEEE Intelligent Vehicles Symposium}, 2003.

\bibitem{lochert_geographic_2005}
C.~Lochert, M.~Mauve, H.~F{\"u}{\ss}ler, and H.~Hartenstein.
\newblock Geographic routing in city scenarios.
\newblock {\em ACM SIGMOBILE mobile computing and communications review}, 9(1):69--72, 2005.

\bibitem{citymesh}
J.~Lynch, Z.~Liu, C.~Li, M.~Ghobadi, and H.~Balakrishnan.
\newblock The case for decentralized fallback networks.
\newblock In {\em Proceedings of the 23rd ACM Workshop on Hot Topics in Networks}, HotNets '24, page 376–384, New York, NY, USA, 2024. Association for Computing Machinery.

\bibitem{ant-dymo}
J.~A.~P. Martins, S.~L. O.~B. Correia, and J.~Celestino.
\newblock Ant-dymo: A bio-inspired algorithm for manets.
\newblock In {\em 2010 17th International Conference on Telecommunications}, 2010.

\bibitem{ant-aodv}
S.~Marwaha, C.~K. Tham, and D.~Srinivasan.
\newblock Mobile agents based routing protocol for mobile ad hoc networks.
\newblock In {\em Global Telecommunications Conference, 2002. GLOBECOM '02. IEEE}, 2002.

\bibitem{wrp}
S.~Murthy and J.~J. Garcia-Luna-Aceves.
\newblock An efficient routing protocol for wireless networks.
\newblock {\em Mob. Netw. Appl.}, 1(2):183–197, 1996.

\bibitem{naumov_evaluation_2006}
V.~Naumov, R.~Baumann, and T.~Gross.
\newblock An evaluation of inter-vehicle ad hoc networks based on realistic vehicular traces.
\newblock In {\em MobiHoc}, 2006.

\bibitem{tbrpf}
R.~Ogier.
\newblock Efficient routing protocols for packet-radio networks based on tree sharing.
\newblock In {\em IEEE Intl. Workshop on Mobile Multimedia Communications (MoMuC))}, 1999.

\bibitem{openwrt}
Openwrt.
\newblock Openwrt.
\newblock \url{https://openwrt.org/}, 2025.

\bibitem{park_1997}
V.~Park and M.~Corson.
\newblock A highly adaptive distributed routing algorithm for mobile wireless networks.
\newblock In {\em Proceedings of INFOCOM '97}, 1997.

\bibitem{tora}
V.~D. Park and M.~S. Corson.
\newblock A highly adaptive distributed routing algorithm for mobile wireless networks.
\newblock In {\em IEEE INFOCOM}, volume~3, pages 1405--1413. IEEE, 1997.

\bibitem{aodv}
C.~Perkins and E.~Royer.
\newblock Ad-hoc on-demand distance vector routing.
\newblock In {\em Proceedings WMCSA'99. Second IEEE Workshop on Mobile Computing Systems and Applications}, 1999.

\bibitem{dsdv}
C.~E. Perkins and P.~Bhagwat.
\newblock {Highly dynamic destination-sequenced distance-vector routing (DSDV) for mobile computers}.
\newblock In {\em ACM SIGCOMM}, 1994.

\bibitem{dymo}
C.~E. Perkins, J.~Dowdell, L.~Steenbrink, and V.~Pritchard.
\newblock Ad hoc on-demand distance vector version 2 (aodvv2) routing.
\newblock Internet-Draft draft-perkins-manet-aodvv2-05, Internet Engineering Task Force, Nov. 2024.
\newblock Work in Progress.

\bibitem{rao_geographic_nodate}
A.~Rao, S.~Ratnasamy, C.~Papadimitriou, S.~Shenker, and I.~Stoica.
\newblock Geographic routing without location information.
\newblock In {\em ACM MobiCom}, 2003.

\bibitem{rao_gpsr-l_2008}
S.~A. Rao, M.~Pai, M.~Boussedjra, and J.~Mouzna.
\newblock Gpsr-l: Greedy perimeter stateless routing with lifetime for vanets.
\newblock In {\em IEEE International Conference on ITS Telecommunications}, 2008.

\bibitem{gabor_ip_forwarding_compress}
G.~R\'{e}tv\'{a}ri, J.~Tapolcai, A.~K\H{o}r\"{o}si, A.~Majd\'{a}n, and Z.~Heszberger.
\newblock Compressing ip forwarding tables: towards entropy bounds and beyond.
\newblock In {\em Proceedings of the ACM SIGCOMM 2013 Conference on SIGCOMM}, 2013.

\bibitem{royer2000implementation}
E.~M. Royer and C.~E. Perkins.
\newblock An implementation study of the aodv routing protocol.
\newblock In {\em IEEE Wireless Communications and Networking Conference. Conference Record}, 2000.

\bibitem{manet_taxonomy}
N.~H. Saeed, M.~F. Abbod, and H.~S. Al-Raweshidy.
\newblock Manet routing protocols taxonomy.
\newblock In {\em 2012 International Conference on Future Communication Networks}, 2012.

\bibitem{localization_error}
K.~Seada, A.~Helmy, and R.~Govindan.
\newblock On the effect of localization errors on geographic face routing in sensor networks.
\newblock In {\em Proceedings of the 3rd International Symposium on Information Processing in Sensor Networks}, 2004.

\bibitem{kanade_-star_2004}
B.-C. Seet, G.~Liu, B.~Lee, C.~Foh, K.~Wong, and K.-K. Lee.
\newblock A-star: A mobile ad hoc routing strategy for metropolis vehicular communications.
\newblock In {\em NETWORKING}, 2004.

\bibitem{shanmuga_gpsr}
B.~Shanmuga~Raja, N.~Prabakaran, and V.~R.~S. Dhulipala.
\newblock Modified gpsr based optimal routing algorithm for reliable communication in wsns.
\newblock In {\em 2011 International Conference on Devices and Communications (ICDeCom)}, 2011.

\bibitem{silva_adaptive_2018}
A.~Silva, K.~N. Reza, and A.~Oliveira.
\newblock An adaptive gpsr routing protocol for vanets.
\newblock In {\em IEEE International Symposium on Wireless Communication Systems (ISWCS)}, 2018.

\bibitem{kranakis_compass}
H.~Singh.
\newblock {\em Compass routing on geometric graphs.}
\newblock University of Ottawa (Canada), 1999.

\bibitem{cedar}
P.~Sinha, R.~Sivakumar, and V.~Bharghavan.
\newblock {CEDAR: a core-extraction distributed ad hoc routing algorithm}.
\newblock In {\em INFOCOM}, 1999.

\bibitem{googleOpenBuilding}
W.~Sirko, E.~A. Brempong, J.~T.~C. Marcos, A.~Annkah, A.~Korme, M.~A. Hassen, K.~Sapkota, T.~Shekel, A.~Diack, S.~Nevo, J.~Hickey, and J.~Quinn.
\newblock High-resolution building and road detection from sentinel-2, 2024.

\bibitem{sklower1991tree}
K.~Sklower.
\newblock A tree-based packet routing table for berkeley unix.
\newblock In {\em USENIX Winter Conf.}, 1991.

\bibitem{srinivasan1999fast}
V.~Srinivasan and G.~Varghese.
\newblock Fast address lookups using controlled prefix expansion.
\newblock {\em ACM Transactions on Computer Systems (TOCS)}, 17(1):1--40, 1999.

\bibitem{batman}
B.~D. Team.
\newblock B.a.t.m.a.n. (better approach to mobile ad-hoc networking).
\newblock \url{https://www.open-mesh.org/projects/open-mesh/wiki}.
\newblock Accessed: 2025-01-06.

\bibitem{champ}
A.~Valera, W.~Seah, and S.~Rao.
\newblock Cooperative packet caching and shortest multipath routing in mobile ad hoc networks.
\newblock In {\em IEEE INFOCOM 2003. Twenty-second Annual Joint Conference of the IEEE Computer and Communications Societies (IEEE Cat. No.03CH37428)}, 2003.

\bibitem{a4lp}
G.~Wang, Y.~Ji, D.~C. Marinescu, and D.~Turgut.
\newblock A routing protocol for power constrained networks with asymmetric links.
\newblock In {\em Proceedings of the 1st ACM International Workshop on Performance Evaluation of Wireless Ad Hoc, Sensor, and Ubiquitous Networks}, 2004.

\bibitem{hopnet}
J.~Wang, E.~Osagie, P.~Thulasiraman, and R.~K. Thulasiram.
\newblock Hopnet: A hybrid ant colony optimization routing algorithm for mobile ad hoc network.
\newblock {\em Ad Hoc Networks}, 7(4):690--705, 2009.

\bibitem{ilar}
N.-C. Wang and S.-M. Wang.
\newblock An efficient location-aided routing protocol for mobile ad hoc networks.
\newblock In {\em 11th International Conference on Parallel and Distributed Systems (ICPADS'05)}, 2005.

\bibitem{aqor}
Q.~Xue and A.~Ganz.
\newblock Ad hoc qos on-demand routing (aqor) in mobile ad hoc networks.
\newblock {\em J. Parallel Distrib. Comput.}, 63(2):154–165, 2003.

\bibitem{yang_compression}
T.~Yang, B.~Yuan, S.~Zhang, T.~Zhang, R.~Duan, Y.~Wang, and B.~Liu.
\newblock Approaching optimal compression with fast update for large scale routing tables.
\newblock In {\em 2012 IEEE 20th International Workshop on Quality of Service}, 2012.

\bibitem{yang_improvement_2018}
X.~Yang, M.~Li, Z.~Qian, and T.~Di.
\newblock Improvement of gpsr protocol in vehicular ad hoc network.
\newblock {\em IEEE Access}, 6:39515--39524, 2018.

\bibitem{mzrp}
X.~Zhang and L.~Jacob.
\newblock Multicast zone routing protocol in mobile ad hoc wireless networks.
\newblock In {\em 28th Annual IEEE International Conference on Local Computer Networks, 2003. LCN '03. Proceedings.}, 2003.

\end{thebibliography}

\end{document}